# Topological design of graphene

*Bo Ni[1,\*], Teng Zhang[2,\*], Jiaoyan Li[1], Xiaoyan Li[3], Huajian Gao[1,†]*

[1]School of Engineering, Brown University, Providence, RI 02912, USA

[2]Department of Mechanical and Aerospace Engineering, Syracuse University, Syracuse, NY 13244, USA

[3]Applied Mechanics Laboratory, Department of Engineering Mechanics, Centre for Advanced Mechanics and Materials, Tsinghua University, Beijing 100084, China

## Abstract

Topological defects (e.g. pentagons, heptagons and pentagon-heptagon pairs) have been widely observed in large scale graphene and have been recognized to play important roles in tailoring the mechanical and physical properties of two-dimensional materials in general. Thanks to intensive studies over the past few years, optimizing properties of graphene through topological design has become a new and promising direction of research. In this chapter, we review some of the recent advances in experimental, computational and theoretical studies on the effects of topological defects on mechanical and physical properties of graphene and applications of topologically designed graphene. The discussions cover out-of-plane effects, inverse problems of designing distributions of topological defects that make a graphene sheet conform to a targeted three-dimensional surface, grain boundary engineering for graphene strength, curved graphene for toughness enhancement and applications in engineering energy materials, multifunctional materials and interactions with biological systems. Despite the rapid developments in experiments and simulations, our understanding on the relations between topological defects and mechanical and physical properties of graphene and other 2D materials is still in its infancy. The intention here is to draw the attention of the research community to some of the open questions in this field.

---

[\*] These authors contributed equally.

[†] Corresponding author. E-mail: Huajian_Gao@brown.edu. Tel: (401)863-2626.

# 1. Introduction

As the very first and most prominent example of two-dimensional (2D) materials, pristine graphene [1] consists of an atomic layer of carbon atoms densely packed in the hexagonal crystal lattice via $sp^2$ covalent bonding. Topological defects in graphene are those induced by re-arrangements of atomic bonds that break the hexagonal symmetry of the 2D lattice. Fundamental units of topological defects in graphene include disclinations [2] (pentagons and heptagons) and dislocations [3] (pairs of pentagon and heptagon), which are disruptions of the rotational and translational symmetry of the lattice, respectively. Grain boundaries [3,4] (GBs) are topological line defects formed between grains with different crystal orientations. Indeed, various forms of topological defects are widely observed in large scale graphene samples fabricated by chemical vapor deposition (CVD) [4-7]. Understanding how they alter the mechanical and physical properties of graphene, including strength [8-11], morphology [2,12,13], toughness [11,14,15], heat conductivity[16], chemical reactivity [17] and electrical properties [18-21], is of great importance in advancing fundamental sciences and applications of 2D materials.

Over the past few years, more and more theoretical studies and experimental observations have shown that mechanical and physical properties of graphene can be tailored by topological defects. For example, molecular dynamics (MD) simulations reveal toughness enhancements in sinusoidal graphene containing periodically distributed disclination quadrupoles [15] and in polycrystalline graphene with well-stitched grain boundaries [14]. Experimental measurements show that the thermal conductivity of polycrystalline graphene decreases dramatically with grain size due to the influence of GBs [22]. MD simulations predict that graphene samples forming a gyroid surface have 300-fold reduction in thermal conductivity due to the presence of topological defects and curvature [23]. Topological defects have also been shown to alter electronic transport behaviors from high transparency to perfect reflection of charge carriers [19]. Recent experimental advances [24-27] have made it increasingly possible to control atomic structure and distribution of topological defects, paving the way for large scale fabrication of "topologically designed" graphene structures and devices.

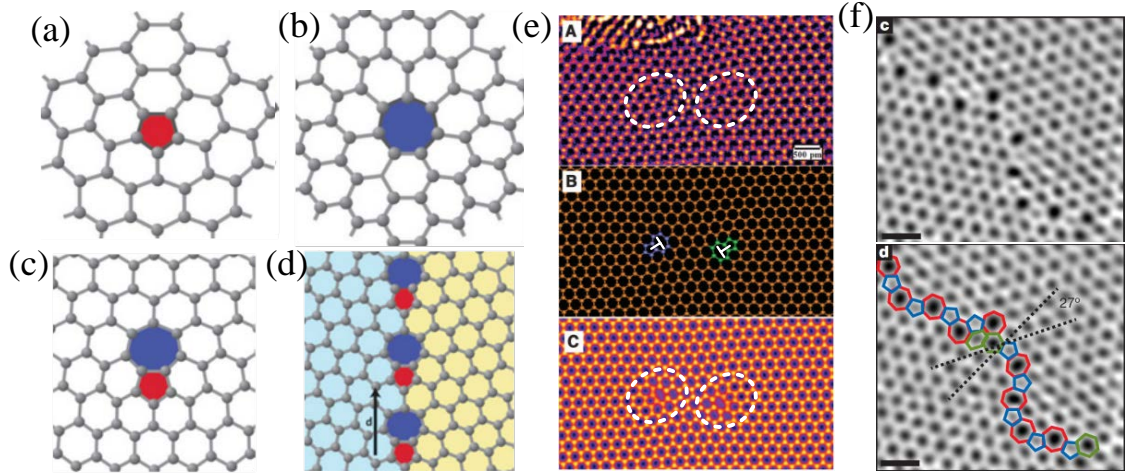

**Fig 1 Topological defects in single-layer graphene. (a)-(d) Schematics of topological defects in graphene, including positive disclination (a), negative disclination (b), dislocation (c) and grain boundary (d) [28]. (e)-(f) Experimentally observed atomic structure of dislocations (e) [29] and grain boundaries (f) [4] in graphene.**

Here, the concept of topological design may be defined as "taking advantages of the cooperative interactions of topological defects, e.g., disclinations, dislocations and grain boundaries, to achieve novel mechanical and physical properties of graphene through design and fabrication of two dimensional lattices with controlled distribution of topological defects." We will focus our discussion to graphene, yet many of the discussions and findings are also applicable to other 2D materials [30,31], such as monolayer hexagonal boron nitride (h-BN) [32,33] and semiconducting transition metal dichalcogenides (TMDCs) $MX_2$ (M = Mo, W; X = S, Se) [34,35]. To avoid the complexity of dangling bonds, we will restrict our discussion to topological defects that do not involve vacancies and free edges, for which relevant papers and reviews including porous graphene [36,37] and kirigami/origami graphene [38-43] can already be found in the literature.

Manipulating nanoscale topological defects to improve the mechanical properties at macroscale is not new and has been widely employed in bulk materials including metals [44,45], ceramics [46] and diamond [47]. For example, it has been well recognized that grain boundaries and twin boundaries play crucial roles in developing novel metallic materials with ultrahigh strength, good ductility and superior fatigue resistance [44,45,48]. Nano-twinned cubic boron nitride [46] and diamond [47] exhibit higher hardness and toughness than their defect-free counterparts. The past successes in topological design of bulk materials provide a solid foundation for the extension of similar concept to 2D materials.

While sharing a number of common features, topological design of graphene exhibits several important distinctions compared to that of bulk materials. First, unlike the bulk materials with abundant slip systems in three dimensions (3D), the migration paths of topological defects in 2D materials are confined within a basal plane [29,49]. This dimensionality restriction largely reduces the accessibility and variability of mechanical behaviors in 2D materials. Second, due to the large differences in the rigidity of in-plane [50] and out-of-plane [51,52] deformations in graphene, the presence of topological defects can trigger substantial out-of-plane deformation, especially in free standing graphene, to minimize its strain energy [3,13]. The resulting 3D geometry will in turn alter mechanical and physical properties like the elastic modulus, strength [9,53], fracture toughness [14,15], adhesion and friction [54], chemical reactivity [17], local density of states [55] and flexoelectricity [56,57]. Third, the flexibility of graphene with respect to the out-of-plane deformation [58] makes its mechanical and physical properties highly sensitive to thermal fluctuations at room temperature [59,60]. As a result, the effective properties of graphene often need to be considered as a consequence of interactions between the intrinsic properties and thermal fluctuations [38,61-64]. Fourth, the atomically thin structure of graphene also poses great challenges in fabrication and post-processing of the material [65-67]. Thus, it should be empathized that the topological design of graphene involves intrinsically nonlinear coupling between many properties including stress, deformation, electricity and chemical reactivity. Addressing these challenges requires a highly interdisciplinary collaboration from multiple communities of researchers such as mechanics, physics, chemistry, material science and nanoengineering. Here, we review some of the recent advances in engineering mechanical and physical properties of graphene through topological design, hoping to draw the attention of various research communities to some of the open questions in this field.

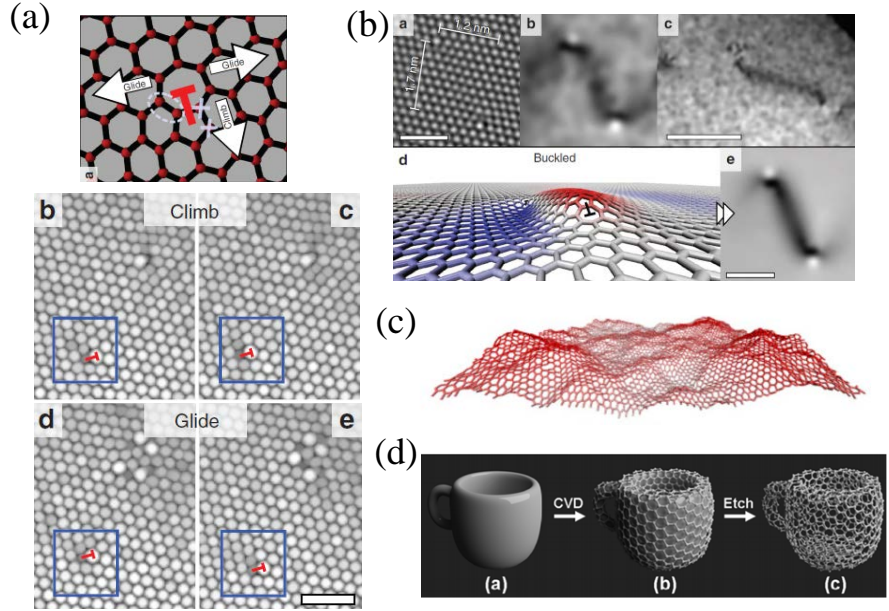

**Fig 2 Unique features in topological design of graphene. (a) Migration of dislocations in single-layer graphene is limited within the lattice plane [49]. (b) The long range out-of-plane deformation triggered by topological defects in graphene in experimental observations and simulations [49]. (c) Atomistic simulations of thermal fluctuations in a free-standing graphene [68]. (d) Schematic of growing graphene on designed curved surfaces via chemical vapor deposition (CVD) methods [69].**

This chapter is organized as follows. In Section 2, we summarize some studies on how to optimize the mechanical and physical properties of graphene through topological design. Section 3 reviews selected applications of topologically designed graphene, examples including applications as novel materials in energy and multifunctional devices. In Section 4, we discuss a few promising techniques to fabricate graphene with deliberately designed topological defects. Finally, some conclusions and outlook remarks will be made in Section 5.

# 2. Topological design for engineering strength, morphology and toughness of graphene

In this section, we will review some of the recent progress on engineering graphene with targeted properties through topological design, with a focus on the mechanical properties (i.e., strength and toughness) and 3D morphology. Examples include how to tailor the strength of graphene by designing grain boundaries in it, how to inversely design the distribution of topological defects to create a 3D single layer graphene with targeted shape, and how to enhance the fracture toughness of graphene via intentionally introduced topological defects.

## *2.1 Tuning strength of graphene via grain boundaries*

For 2D materials, a GB is a one-dimensional chain of edge dislocations. Thus, a GB can be regarded as a simple linear array of topological defects. For single-layered 2D materials, there is only edge dislocation and no screw dislocation, because all dislocations and their projections are located in the basal plane. The edge dislocation in graphene is described by the Burger vector **b**, a topological invariant. In 2D materials, an array of edge dislocations constitutes a tilting GB, usually described by a misorientation angle or tilting angle $\theta$, that separates two grains with different crystal orientations. Recent high-resolution transmission electron microscopy (HRTEM) observations showed the detailed atomic structures of some GBs in graphene [4,70,71], hexagonal boron nitride (h-BN) [72] and TMDCs [73]. GBs in 2D materials usually form during growth. For example, single-layer graphene can be synthesized by CVD [4,70,71] on metal substrates with a certain crystalline orientation. During synthesis, independent grains simultaneously nucleate at different points on the metal surface, and the misfit between graphene and metal leads to different lattice orientations in different grains. When two grains with different orientations meet, a line defect, i.e. a GB, forms along the interface.

Before we discuss how we can tune the strength of graphene with GBs, let us quickly review the atomic structures and energies of GBs in graphene. In the single-layer graphene, GBs are generally made of two types of edge dislocations, one type with Burgers vector **b** = (1,0) or (0,1) consisting of a pair of neighboring pentagon and heptagon (Fig 3a) and another with the distance between pentagon and heptagon increasing by one lattice length, resulting in Burgers vector of **b** = (1,1), as shown in Fig 3b. Periodically aligning edge dislocations along a certain direction leads to a GB. Recent first-principles calculations revealed the atomic structures of some energetically favorable GBs in graphene consistent with HRTEM observations and also provided a diagram of GB energies per unit length $\gamma$ as a function of the tilting angle $\theta$ for various GB structures (Fig 3c) [3]. When the GBs are confined in a 2D plane, their energies in the small-angle regime ($\theta < 10°$) can be described by the Read-Shockley equation [3]

$$\gamma = \frac{\mu b \theta'}{4\pi(1-\nu)}\left(1 - \ln\theta' + \ln\frac{b}{2\pi r_0}\right) \tag{1}$$

where $\mu$ is the shear modulus, $\nu$ is the Poisson's ratio, $b$ is the magnitude of Burgers vector, and $r_0$ is the radius of dislocation core. In Eq. (1), $\theta' = \theta$ or $\theta' = \pi/3 - \theta$ for armchair and zigzag GBs, respectively. If there are no planar constraints, the GB will exhibit a buckled shape due to the out-of-plane deformation associated with dislocations. Such buckling reduces the energy of GBs (open data points in Fig 3c), making them more stable. There exist two particularly stable large-angle GBs with $\theta$ = 21.8° and 32.3°, which hold the lowest energies in the $\theta < 21.8°$ and $\theta > 32.3°$ regimes, respectively. For buckled GBs with $\theta < 3.5°$, the GB energies exhibit a linear dependence on the tilting angle, yielding the following expression [3],

$$\gamma = \frac{E_f \theta'}{b} \tag{2}$$

where $E_f$ represents the formation energy of a dislocation. The fitting in Fig 3c gave $E_f$ as 7.5 eV [3].

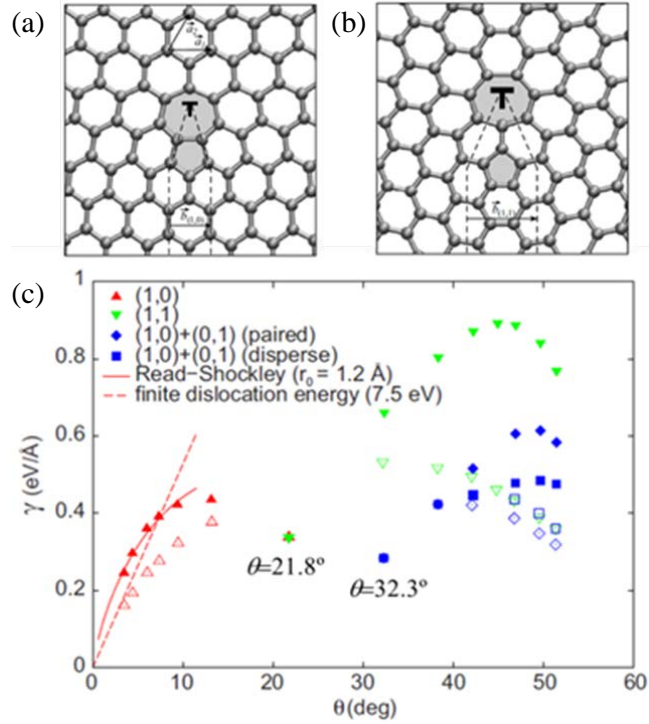

Fig 3 Atomic structures of GB dislocations and GB energies in graphene. (a-b) Atomic structures of dislocations with b = (1,0) and b = (1,1). (c) GB energies per unit length as a function of the tilting angle. Solid and hollow data points are for flat and buckled graphene, respectively. Solid curve is a fitting curve with dislocation core radius of $r_0$ = 0.12 nm based on the Read-Shockley equation, while the dashed curve reflects the asymptotic linear expression with $E_f$ =7.5 eV based on Eq. (2) for buckled GBs. Figures reprinted with permission from ref. [3].

The effects of GBs on mechanical strength of polycrystalline graphene have been widely explored by both experiments and simulations. Lee et al. [50] have experimentally studied the mechanical properties of CVD-graphene films with different grain sizes by combining TEM structural characterization with nanoindentation. They found that the elastic stiffness of CVD-graphene is close to that of pristine graphene while the mechanical strength is slightly reduced. Rassol et al. [10] performed nanoindentation on a bi-crystalline graphene and found that the mechanical strength of GBs with large mismatch angles are larger than that of GBs with low mismatch angles. This observed misorientation dependence of GB strength corroborates predictions from atomistic simulations [8,9,74]. Using atomistic calculations, Grantab et al. [8] showed that the large-angle GBs are stronger. Wei et al. [9] combined continuum modeling and atomistic simulations to study how defects in GB interact; Their results emphasized that it is not only the density of defects that affects the mechanical properties, but the detailed arrangements of the defects are also important to GB strength. Later on, further efforts [75] have extended the studies from symmetric to asymmetric GBs. Besides straight GBs, Zhang et al. [74] investigated sinuous GBs which are frequently observed in experiments; They concluded that the sinuous GBs can be more energetically favorable than straight ones and have improved mechanical properties.

Beyond the above studies on the properties of GBs as line defects, designing networks of GBs [76-78] has also attracted increasing attention as a promising way to control and/or tune physical properties of 2D materials. Recent advances of fabrication techniques showed a great potential to control GBs in polycrystalline graphene during the growth process. For example, seed-assisted growth has been successfully carried out in experiments by suppressing random nucleated islands via an array of pre-patterned seeds [79-81]. Besides the locations of nucleated islands, the shape, orientation, and edge geometry of CVD graphene domains are also controllable by the crystallographic orientations of copper substrates [79,82]. Based on a phase field crystal model [83], Li et al. [84] numerically simulated the dynamic formation of GBs in CVD graphene and demonstrated possible routes of engineering GBs by controlling grain orientations in pre-patterned growth seeds. This study provided a theoretical platform to explore the potential rational design of GBs using pre-patterned growth seeds. In a simple geometrical model to understand the dynamic coalescence of growing seeds, the direction and misorientation angle of a GB are determined from the geometries of polygonal graphene flakes [85]. Integrating this geometrical rule and the phase field crystal model, Li et al. [84] demonstrated that starting with random seeds, serpentine GBs and triple junctions (TJs) are likely to appear, which is consistent with experimental observations in CVD-grown polycrystalline graphene. Extensive studies have shown that the strength of polycrystalline graphene is not only dependent on the grain size but also highly sensitive to the detailed distribution of topological defects in the GB network, such as TJs and vacancies [78,86-88]. As a prominent example of engineering GBs with seed-assisted growth, a design of triple-junction-free polycrystalline graphene [84] has been proposed (illustrated in Fig 4) where GBs are of 30º mis-orientation angles, leading to enhanced, grain-size-insensitive mechanical strength that defies the reported Hall-Petch type relation for polycrystalline graphene [89].

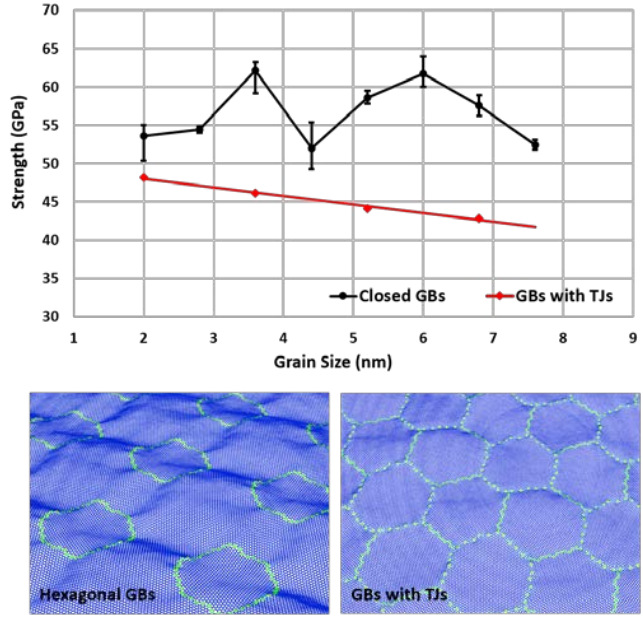

Fig 4 Mechanical Strength of TJ-free graphene with hexagonal GB loops and polycrystalline graphene with TJs under biaxial tension for grain sizes from 2nm to 10nm. Figures reprinted with permission from ref. [84].

Besides graphene, it may be possible to generalize the concept of GB engineering for strength to other 2D materials. For this purpose, it might be interesting to compare similarities and differences in the fundamental structures and energies of GBs in other two typical 2D materials, h-BN and TMDC $MS_2$ (M = Mo or W) with those in graphene. In the single-layer h-BN, apart from the dislocation with pentagon-heptagon pair and **b** = (1,0), a new dislocation structure with square-octagon pair and **b** = (1,1) has been predicted by first-principle calculation [90]. In h-BN, a pentagon-heptagon pair contains energetically unfavorable homo-elemental bonds B-B or N-N, while a square-octagon pair involves the hetero-elemental bond B-N and is free of homo-elemental bonds. As a result, the dislocation of square-octagon pair has lower energy than that of pentagon-heptagon pair [90]. The stabilization of dislocation with square-octagon pair is to some extent associated with its out-of-plane buckling [90]. According to the mirror symmetry and hetero-elemental composition, GBs in h-BN can be classified into two types, i.e. symmetric armchair GBs (A-GBs) and asymmetric zigzag GBs (Z-GBs), as shown in Fig 5a. The symmetric A-GB consists of dislocations with pentagon-heptagon pairs, while the asymmetric Z-GB is composed of dislocations with square-octagon pairs [90] (Fig 5a). The symmetric A-GB has been observed by HRTEM [72]. Due to the elemental polarity (either B- or N-rich) along a GB, the symmetric A-GB is capable of carrying net charges [90], which suggests possible applications in electronic and optical devices. Fig 5b presents the GB energy per unit length as a function of the tilting angle [90]. It is seen that the energies of GBs composed of square-octagon pairs are always lower than those of pentagon-heptagon pairs. The dislocations aligned along a GB induce out-of-plane buckling. The asymmetric Z-GBs exhibit more buckling compared to the symmetric A-GBs, which helps to reduce the energies of the Z-GBs.

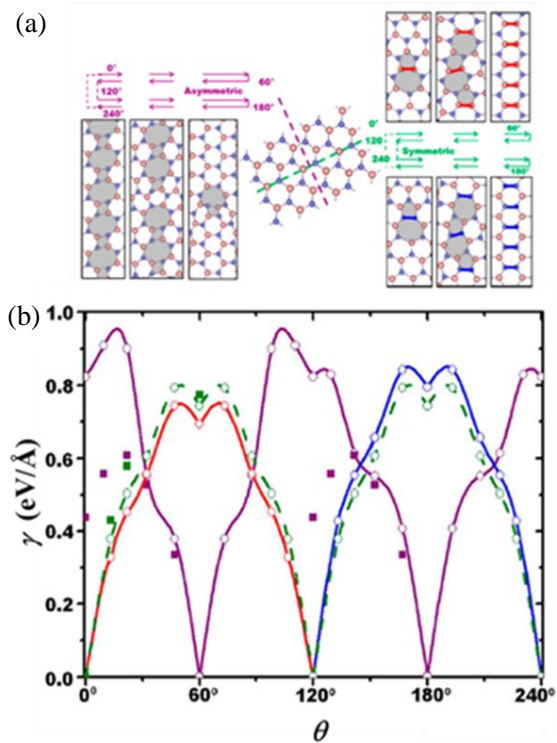

Fig 5 Atomic structures and energies of GBs in h-BN. (a) Atomic structures of GBs. The middle picture shows a perfect lattice. The symmetric A-GBs (right figures) and asymmetric Z-GBs (left figures) are generated by rotation of two grains with respect to green and purple lines, respectively. (b) GB energies per unit length as a function of tilting angle. Scatted data points are from first-principle calculations and are connected by guide lines. Hollow circles are for GBs constituted by dislocations with pentagon-heptagon pairs, while solid squares are for GBs constituted of dislocations with square-octagon pairs. Purple for Z-GBs, red for B-rich A-GBs, blue for N-rich A-GBs, and green for the average energy of B-rich GBs and their N-rich analogs. Figures reprinted with permission from ref. [90].

A sheet of TMDC $MS_2$ (M=Mo or W) is a sandwiched structure containing a midplane of metal atoms and two layers of sulfur atoms triangularly packed in their respective planes. Such three-atomic-layer structure makes dislocation structure more complicated compared with mono-atomic-layer graphene and h-BN. Recent first-principle calculations [91] predicted that there exist three types of edge dislocations in TMDC $MS_2$ which extend through the triatom layers and form concave dreidel-shaped polyhedra [91]. In the planar view, the three types of dislocations are constituted by pentagon-heptagon pairs with M-M bonds, pentagon-heptagon pairs with S-S bonds and square-octagon pairs with M-S bonds [91], with Burgers vectors of (1,0), (0,1) and (1,1), respectively. Due to the local-chemical energy, the dislocation cores in the TMDC $MS_2$ can reconstruct or react with the point defects [91]. For example, an isolated dislocation with square-octagon pair is unstable and can split into two dislocations (one with **b** = (1,0) and the other with **b** = (0,1)) via exothermic reconstruction [91]. Similar to h-BN, GBs in TMDC $MS_2$ can be classified into two types: A-GBs and Z-GBs (Fig 6). But these GB structures are more complex than those of h-BN. Recent HRTEM observations [73] showed that the Z-GBs with square-octagon pairs are dominant in the samples fabricated by CVD. Figure 3 shows the GB energies per unit of length as function of tilting angles [91]. In the large-angle regime, the GB energies have a large variation due to the reconstructions of dislocation cores or their reaction with point defects.

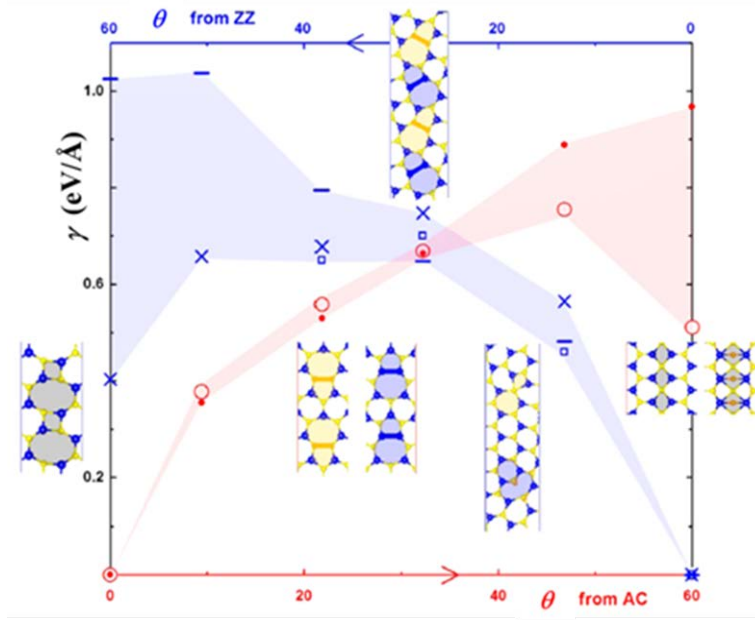

**Fig 6** Atomic structures and energies per unit length of GBs as function of tilting angle in TDMC $MS_2$. Atomic structures of some specific GBs are shown in the insets. GB energies per unit length change with the tilting angles along armchair (AC) and zigzag (ZZ) directions. Red solid and open circles correspond to A-GBs composed of pentagon-heptagon pairs and rhomb-hexagon plus hexagon-octagon pairs, respectively. Blue dashes, crosses and open squares correspond to Z-GBs composed of pentagon-heptagon pairs, rhomb-hexagon plus hexagon-octagon pairs and square-octagon pairs, respectively. Shaded areas show the energy range due to reconstruction of dislocation cores. Figure reprinted with permission from ref. [91].

## 2.2 Topological design for 3D shapes of graphene

Since graphene is a highly flexible atomic thin crystal membrane, it will adopt a 3D configuration to release strain energy induced by topological defects. As shown in Fig 7, global 3D configurations can be formed by introducing a single disclination, such as a cone shape for a pentagon and a saddle surface for a heptagon. Even an isolated dislocation can cause substantial out-of-plane deformation [13]. It has been widely recognized that the shape of graphene plays a crucial role in determining its mechanical [15,92], thermal [23], chemical [93] and physical properties [94-96]. If we can design graphene with arbitrary shapes by deliberately controlling the topological defects in it, there will be tremendous opportunities to tailor its properties for specific applications. For example, the super strong yet brittle pristine graphene [97] may not be the best candidate for reinforcing light-weight, strong and tough composites, where an alternative structure with designed topological defects to achieve balance properties among strength, toughness and interfacial adhesion may be more desirable.

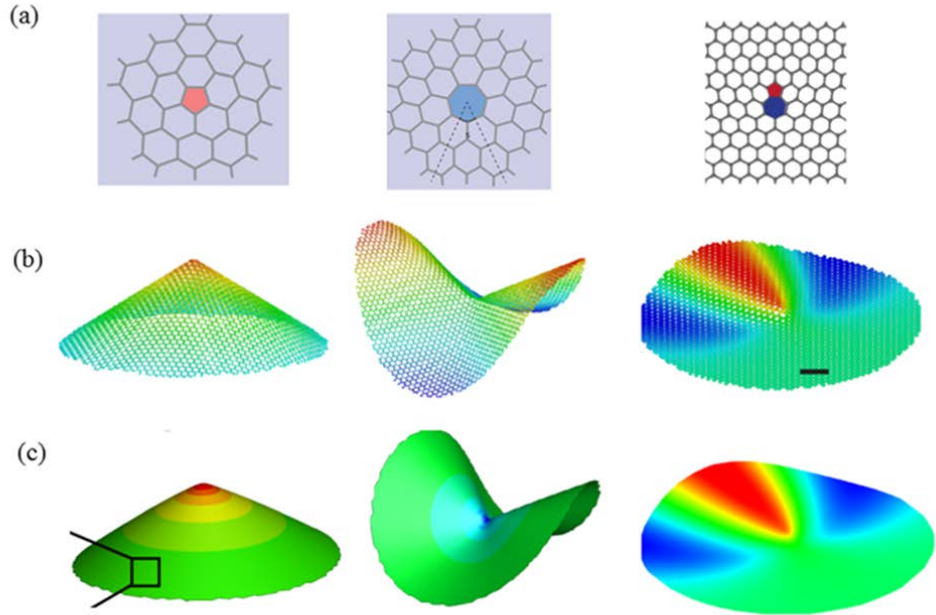

**Fig 7 3D curved shapes induced by elementary topological defects in graphene [13]. (a) Atomic structures of positive and negative disclination and edge dislocation in graphene. (b) 3D configurations from MD simulations. (c) 3D configurations from continuum model.**

Achieving an inverse design of 3D curved graphene with a targeted shape is a very challenging task, as we need to search for the number, type and location of corresponding topological defects. The first challenge comes from highly nonlinear interactions between the topological defects and 3D shapes of graphene in the forward analysis for a given defect distribution [13,98]. The second challenge is due to the multiple time scales involved in directly optimizing the carbon atom positions, which is generally at the level of seconds, which far exceeds the current capability of molecular dynamics (nanoseconds) [99,100]. Other techniques, like geometrical methods [101-103] and Monte Carlo simulations [104,105], may provide a path to bridging the time scales, but still require a large amount of computation efforts, especially for large graphene structures. The third challenge stems from the fabrication techniques for realizing topologically designed graphene. In this section, we will review some of the recent developments in predicting 3D curved graphene with topological defects via continuum models and inverse design of 3D curved graphene through phase field crystal methods. It remains an open question how to fabricate 3D curved graphene, and we will review some promising techniques in Section 3.

Studies of buckling of plates with defects can be traced back to the 1960s, when Mitchell and Head investigated the critical buckling condition of a plate with a central dislocation based on an energy method [106]. In 1988, Seung and Nelson [107] derived a generalized von Karman equation for thin elastic sheets with various topological defects and validated the theoretical predictions of shape and energy via a triangular lattice model. Zubov [108-110] conducted a series of studies on thin shells and plates with topological defects and showed that the problem of a thin shell with defects can be linked to its dual problem of a thin shell with external loading [110]. Chen and Chrzan [98] formulated a continuum theory for dislocations in graphene by modeling dislocations as topological constraints and minimizing the total strain energy in the Fourier space, which was shown to accurately capture the self-energy of periodical dislocation dipoles in a graphene sheet with out-of-plane deformation compared with MD simulations. Zhang et al. [13] developed a continuum model of topological defects in graphene in terms of a classical von Karman equation with eigenstrain field based on a mathematical analogy between topological defects and incompatible growth metric field. The model proposed by Zhang et al. [13] successfully captured the global wrinkling profiles and atomic scale wrinkles near disclination/dislocation cores, with much higher efficiency compared to full atom MD simulations [13].

In the generalized von Karman model of a 2D lattice with topological defects [107], the out-of-plane deformation $w$ and Airy stress function $\Phi$ are expressed as

$$B\nabla^4 w = [w, \Phi]$$
$$\nabla^4 \Phi = -S\left[\kappa_G - \sum_{i=1}^{N} s_i \delta(\mathbf{r} - \mathbf{r}_i)\right]$$
(3)

where $S = Eh$ denotes the in-plane stretching stiffness, $s_i\delta(\mathbf{r}\text{-}\mathbf{r}_i)$ represents the $i_{th}$ disclination at position $\mathbf{r}_i$ with strength $s_i$, $\nabla^4$ is the bi-harmonic operator, and $[f, g]=f_{,11}g_{,22}+ f_{,22}g_{,11} -2f_{,12}g_{,12}$. Interestingly, a similar governing equation has also been derived for inhomogeneous growth of a thin film [111,112],

$$B\nabla^4 w = [w, \Phi]$$
$$\nabla^4 \Phi = -S\left[\kappa_G + \lambda_g\right] \qquad (4)$$

where $\lambda_g = \varepsilon^g_{11,22} + \varepsilon^g_{22,11} - 2\varepsilon^g_{12,12}$ is the incompatibility metric due to in-plane growth or swelling. It is noted that the two sets of equations (i.e. Eqs. (3) and (4)) are identical if one sets $\lambda_g = -\sum_{i=1}^{N} s_i\delta(\mathbf{r}-\mathbf{r}_i)$. In principle, $\varepsilon^g_{11}, \varepsilon^g_{22}$ and $\varepsilon^g_{12}$ can be chosen independently, as they do not necessarily satisfy the incompatible conditions. One possible choice is $\varepsilon^g_{12} = 0$, $\varepsilon^g_{11} = \varepsilon^g_{22} = \varepsilon^g$, which will lead to a Poisson's equation for $\varepsilon^g$,

$$\nabla^2 \varepsilon^g = -\sum_{i=1}^{N} s_i\delta(\mathbf{r}-\mathbf{r}_i) \qquad (5)$$

The fundamental solution of Eq. (5) in an infinite domain can be written as,

$$\varepsilon^g = -\sum_{i=1}^{N} \frac{s_i}{2\pi}\log|\mathbf{r}-\mathbf{r}_i| + \text{constant} \qquad (6)$$

where the constant value should be determined from boundary conditions.

Topological defects in graphene can be represented by the following growth strain field in a perfect continuum film [13],

$$\varepsilon^g_{12} = 0, \ \varepsilon^g_{11} = \varepsilon^g_{22} = -\sum_{i=1}^{N} \frac{s_i}{2\pi}\log|\mathbf{r}-\mathbf{r}_i| + \text{constant} \qquad (7)$$

A Gaussian function $\frac{1}{\pi r_c^2}\exp\left[-\frac{(\mathbf{r}-\mathbf{r}_i)^2}{r_c^2}\right]$ with an intrinsic length scale $r_c$ can be used to replace $\delta(\mathbf{r}-\mathbf{r}_i)$ to eliminate the singularity at the center of defects, which then modify the solution to Eq. (5) as,

$$\varepsilon^g = -\frac{s_i}{2\pi}\left[\log(|\mathbf{r}-\mathbf{r}_i|) - \frac{1}{2}\mathrm{Ei}\left(-\frac{(\mathbf{r}-\mathbf{r}_i)^2}{r_c^2}\right)\right] + \text{constant} \tag{8}$$

where $Ei(\mathrm{x})$ is the exponential integral.

Zhang et al. [13] implemented the above continuum model into a triangle lattice model and simulated the wrinkle patterns of graphene with isolated pentagons (negative disclination), heptagons (positive disclination) and pentagon-heptagon pairs (dislocation). The 3D configurations predicted by the continuum model are in good agreement with full atom simulations based on the AIREBO potential [113] (Fig 8). Inspired by the significant 3D deformation of graphene with simple topological defects, it was further shown from continuum and atomistic simulations that periodically distributed disclination quadrupoles lead to a sinusoidal graphene ruga[1] (Fig 8 a), and an array of dislocations on a cylindrical graphene can deform the carbon tube into a catenoid graphene funnel (Fig 8 b).

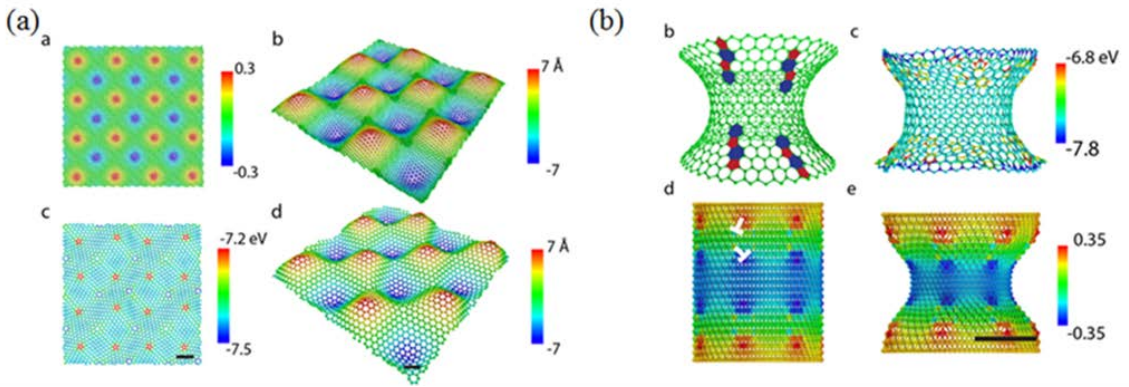

**Fig 8 Sinusoidal graphene and catenoid graphene funnel achieved via topological design [13]. (a) A sinusoidal graphene induced by a periodic array of disclinations from continuum and atomistic simulations. (b) A catenoid graphene funnel from atomic simulations and continuum modeling.**

---

[1] The Latin word *ruga* is used to refer a large-amplitude state of wrinkles, creases, ridges or folds [114].

These successes suggested the possibility to design arbitrarily curved graphene with topological defects. However, a direct search of atomic positions for curved graphene from molecular dynamics is prohibited by the huge time scale gap between atom diffusion in graphene (seconds to hours) and the typical time scale associated with MD simulations (~ nano seconds) [99,100]. Other studies have attempted to employ geometrical methods [101-103] and Monte Carlo simulations [104,105] to search for equilibrium positions of carbon atoms on a curved surface. Zhang et al. [15] developed a general design methodology by combining a phase field crystal (PFC) method [83] and MD simulations. The PFC method [83] can describe the defect motions in crystalline structures on both flat and curved configurations through over-damped conservative (diffusive) dynamics [115], which is a key to bringing realistic time scale in simulations with atomistic spatial resolutions.

The PFC model can be defined through the following free energy functional [83],

$$F = \int \left[ \frac{\phi}{2} \left( -\varepsilon + (1+\Delta)^2 \right) \phi + \frac{1}{4} \phi^4 \right] d\mathbf{x} , \qquad (9)$$

where $\Delta = \frac{\partial^2}{\partial x^2} + \frac{\partial^2}{\partial y^2}$ is the Laplace operator in 2D, $\phi$ the reduced density and $\varepsilon$ the reduced temperature. The governing equation for the dynamics of density evolution can be defined as,

$$\frac{\partial \phi}{\partial t} = \Delta \left\{ \left( -\varepsilon + (1+\Delta)^2 \right) \phi + \phi^3 \right\} . \qquad (10)$$

To handle complex geometry, Eq. (10) can be solved using finite element method (FEM) by re-writing it in the following form,

$$\begin{aligned} \frac{\partial \phi}{\partial t} &= \nabla^2 \mu \\ \mu &= (-\varepsilon + 1)\phi + 2u + \nabla^2 u + \phi^3 , \\ u &= \nabla^2 \phi \end{aligned} \qquad (11)$$

where two new variables $(\mu, u)$ are introduced to convert the order of the sixth order partial differential equation (PDE) to a set of second order PDEs [116]. Eq. (11) can be implemented in the standard FEM framework and solved efficiently by leveraging open source software packages like FEniCS [116].

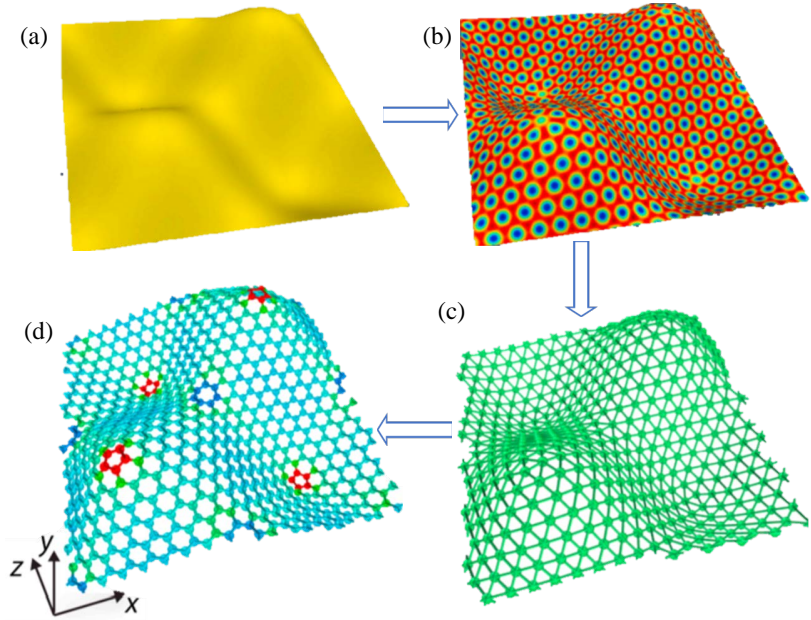

**Fig 9 A general methodology to design an arbitrary 3D curved graphene structure through controlled distributions of topological defects via a combination of phase field crystal (PFC) and atomistic methods [15]. (a) The targeted curved surface. (b) A continuum triangular pattern of density waves on the targeted curved surface generated by PFC. (c) A discrete triangular lattice network from the continuum density waves. (d) The full-atom structure generated by Voronoi construction from the triangular network, followed by equilibration through MD simulations.**

Taking the sinusoidal graphene as an example (Fig 9), the design methodology can be summarized as follows. First, PFC simulation is applied on the targeted curved manifold (Fig 9a), whose solution leads to an equilibrium triangular pattern of continuum density waves corresponding to a minimum energy state (Fig 9b). Second, a discrete triangular lattice network is obtained by identifying wave crests of the continuum triangular pattern of density as particles (Fig 9c). Third, a full-atom graphene structure is generated via Voronoi construction on the triangular lattice. Finally, a thermodynamically stable structure (Fig 9d) is achieved through MD equilibration at a finite temperature. Zhang et al. [15] showed that the predicted atomic structure of sinusoidal graphene agrees well with that from previous Monte Carlo simulations [105] of particle patterns confined on a sinusoidal surface. The new method is efficient and flexible enough to handle complex geometries and can be used to design graphene with targeted shapes through controlled distributions of topological defects. The designed sinusoidal graphene exhibits interesting properties such as enhanced toughness [15], tunable friction [54] and even negative Poisson's ratio [117].

**2.3 *Topological design for toughening graphene***

Toughness is defined as the elastic energy released per unit area associated with advancement of a crack [118], thereby characterizing the resistance of a material to fracture in the presence of crack-like flaws. Sufficiently high toughness is important to ensure mechanical reliability of graphene for applications in practical devices/systems. However, experimental studies have shown that pristine graphene has a fracture energy as low as 16 *J/m$^2$* [87], close to that of an ideally brittle solid, even though it is the strongest materials with a Young's modulus of 1 *TPa* and a strength of 130 *GPa* [50]. During the design, large scale fabrication and postprocessing operations of real devices and systems with graphene (e.g., CVD growth [5,119,120], transfer between different substrates [121-123], patterning and etching [124-126]), various forms of geometrical flaws (e.g., holes, notches, and cracks) may be introduced. This makes the actual failure strength of graphene determined by its ability to resist crack growth. Moreover, when corrosive species, like water vapor, are present in the working environment, stress corrosion cracking could further reduce the fracture resistance of material [127]. Thus, the inevitable flaws and corrosive environments make fracture one of the most prominent concerns in large scale applications of graphene and are calling for efforts to explore effective methods to toughen graphene and other 2D materials [128-135]. In this section, we will review some of the recent progresses on toughening graphene through topological design.

In bulk materials, it is well known that topological defects like dislocations and grain boundaries play important roles in tuning deformation mechanisms and fracture behaviors of many types of materials, including metal [44,45,136,137], ceramics [46] and diamond [47] (Fig. 10). For instance, nanoscale engineering of grain boundaries and twin boundaries [44,45] have been widely employed to design superior metals with high strength and toughness (Fig 10a). The fracture toughness of ceramics (like $Al_2O_3$, Si) [138,139] shows a more than two-fold enhancement by creating a high density of tangled dislocations in the sub-surface region (Fig 10b). Nano-twined cubic boron nitride [46] and diamond [47] exhibit higher hardness and toughness than their defect-free counterparts (Fig 10c). The success of topological design in toughening bulk materials poses the question whether graphene and 2D materials in general can be toughened by designed topological defects. Recent studies in this direction have unveiled a number of toughening mechanisms induced by topological defects, including stress shielding, crack branching, atomic chain bridging and stress reduction due to 3D geometry and nano-crack shielding.

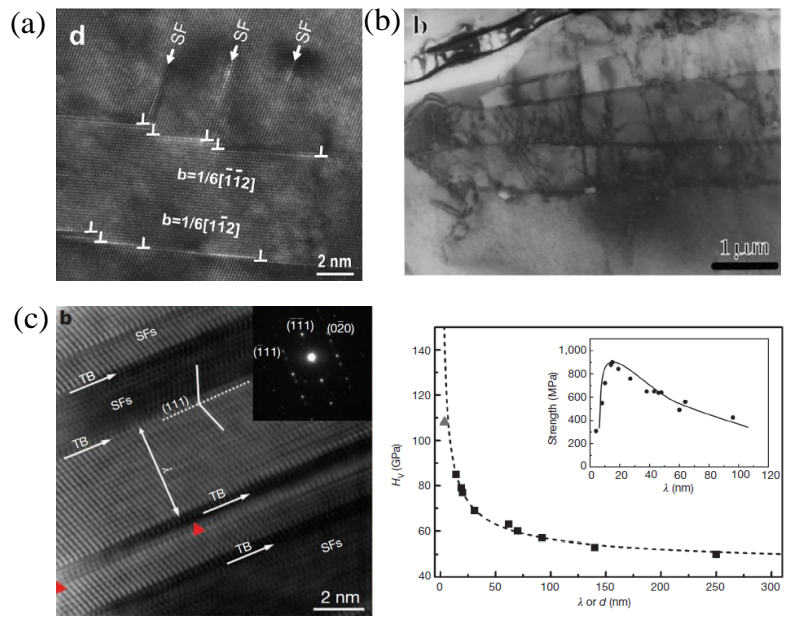

Fig 10 Topological defect design in bulk materials with enhanced mechanical properties. (a) The interaction of dislocations with nanoscale twin boundaries in pure Cu deformed in tension [45]. (b) Tangled dislocations form well-defined sub-boundaries beneath the surface of bulk Al2O3 after shot blasting and annealing [138]. (c) HRTEM image of twin boundaries in nanotwinned cubic boron nitride and the hardness of nt-cBN as a function of averaged grain size, *d* or twin thickness *λ* [46].

The interaction between crack tip and stress induced by topological defect results in stress shielding, which is an important toughening mechanism in topologically designed graphene. Recent theoretical and numerical studies have shown that topological defects can alter the crack tip stress field and induce effective toughness enhancement [140-146]. For example, via MD simulations, it has been demonstrated that the stress resulting from individual topological defects like dislocations (a pentagon-heptagon pair) (Fig 11a) [140], Stone-Thrower-Wales defects (5-7-5-7 rings) (Fig 11b) [141,142,144] and 5-8-5 defects (Fig 11c) [143] can alter the crack tip stress intensity factor. Combining MD simulations and continuum modeling, Meng et al. [140] showed that dislocation shielding in graphene agrees well with the prediction of continuum linear elastic fracture mechanics (LEFM) (Fig 11a). By arranging topological defects into regular or irregular grain boundaries, researchers [14,145,146] have studied more complicated interactions between topological defects and crack tip during propagation (Fig 11d-f). In addition to dislocation shielding, other toughening mechanisms including crack branching and atomic chain bridging can be activated during the crack propagation process. For instance, through MD simulations, Jung et al. [14] showed that the weak points within pentagon-heptagon defects can break near a crack tip resulting in crack branching and atomic chain bridging in polycrystalline graphene samples (Fig 11e-f). Combining these toughening mechanisms together, a 50% enhancement in fracture toughness was reported in a polycrystalline graphene sample with well-stitched randomly distributed grain boundaries [14].

Compared with bulk materials, the interaction between crack tip and topological defects in graphene has several unique features. For atomic thin membranes like graphene, it is essential to consider non-local coupling between out-of-plane deformation and topological defects. On one hand, it has been demonstrated that the residual stress resulting from the in-plane lattice distortion of a topological defect can be partially released through out-of-plane deformation [3,13]. This 3D relaxation tends to weaken the effect of the residual stress of topological defects on a crack tip. On the other hand, the out-of-plane deformation induced by topological defects also changes the sample shape in the global level [13], including the region where a crack tip resides, which may reduce the effective stress intensity near the crack tip and toughen the 2D material. This non-local interaction has been demonstrated by the MD simulations of Jung et al. [14], where the toughening effect of grain boundaries diminishes when the out-of-plane deformation is constrained even though the distribution of topological defects remains the same (Fig 11f). It is the competition of these effects that determines the overall toughness enchantment.

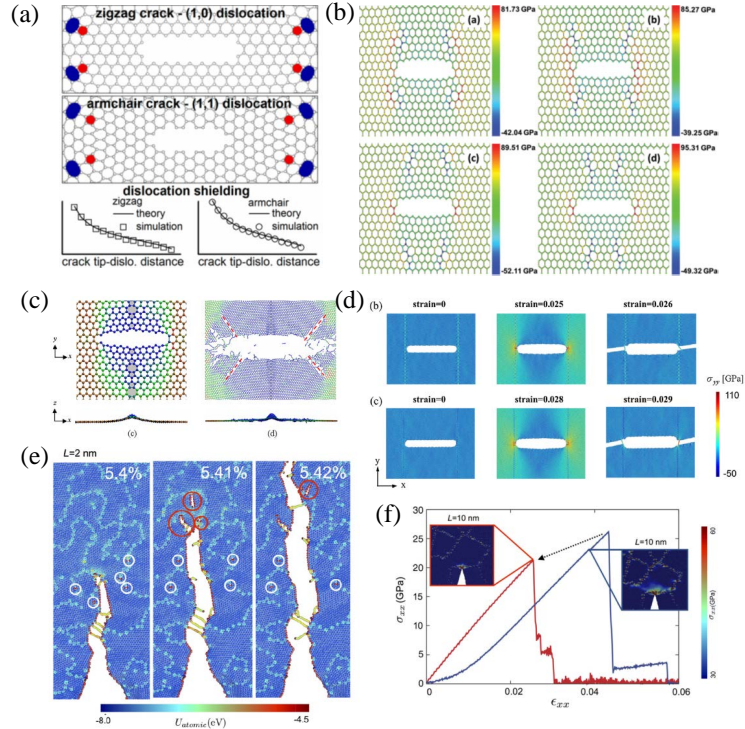

**Fig 11 Interactions between cracks and topological defects in graphene. (a) Dislocation shielding of crack tips [140]. (b) Stress distribution in a graphene sheet with an embedded crack and STW defects placed at various positions [141]. (c) Stress distribution and failure process in a graphene sheet with a finite crack and 5-8-5 defects [143]. (d) Fracture of armchair-oriented bi-crystalline graphene with a finite crack and symmetric tilt GBs of different misorientation angles [145]. (e) Atomic energy distribution during crack propagation in a polycrystalline graphene with irregular grain shapes [14]. (f) Stress-strain curves of a cracked polycrystalline graphene of irregular grain shapes with (blue curve) and without (red curve) out-of-plane relaxation [14].**

The interaction between cracks and 3D curved geometry of graphene results in another group of toughening mechanisms. As discussed in Section 2.2, a 3D curved sinusoidal graphene with a wavelength of 4 *nm* and an out-of-plane amplitude of 0.75 *nm* can be designed by periodically arranging disclination quadrupoles in graphene [15]. The mode I fracture toughness of this sinusoidal graphene is about 25.0 $J/m^2$ based on MD simulations, nearly twice that of pristine graphene (Fig 12a-b). It was found that the sinusoidal geometry and distributed defects give rise to stress reduction near the crack tip, nano-crack initiation at the defected sites and atomic scale crack bridging. As depicted in Fig 12c, the non-planar sinusoidal geometry leads to a non-uniform stress field in the sample and a moving crack tip can be trapped in less stressed regions. In addition, the topological defects within this non-uniform deformation field tend to fail at bonds with high pre-stress (e.g. bonds shared by heptagon and hexagon rings) [9,75], leading to discrete rupture events ahead of a trapped crack tip, forming a nano-crack that shields the main crack. The deformation of the material connection between the nano-crack and main crack results in an atomic scale chain bridging mechanism on crack advance.

The above example of interaction between crack and topologically designed 3D shape of graphene illustrates that fracture in topologically designed 2D materials needs to be treated as a fully 3D problem with strong nonlinearity. The sinusoidal graphene is a consequence of the out-of-plane relaxation of lattice frustration due to the periodically distributed disclination quadrupoles, which can only be understood by modeling fracture along a 2D topological manifold in a 3D space. In such a 3D model of fracture in 2D materials, there exists strong nonlinear coupling between topological defects, out-of-plane geometry and deformation, and crack propagation behaviors. By studying the fracture behavior of rubber sheets draped on curved surfaces, Mitchell et al. [147] demonstrated that surface curvature alone could stimulate or suppress crack propagation via the curvature-induced stress at the continuum level (Fig 12d). At the atomic level, the case of sinusoidal graphene shows that, apart from the curvature-induced stress, topological defects provide sites of nanocrack nucleation ahead of the main crack, resulting in additional toughness enhancement.

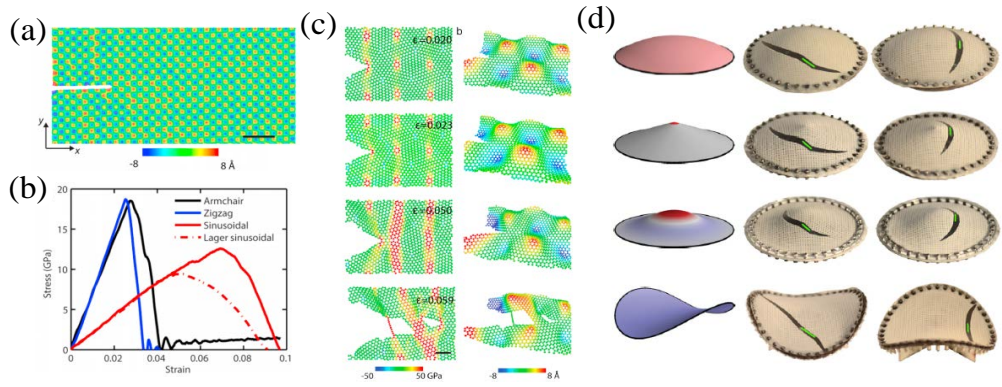

**Fig 12 Crack propagation behaviors in sinusoidal graphene and thin rubber sheets. (a) A nanostrip of sinusoidal graphene with an edge crack [15]. (b) Stress-strain curves of pristine and sinusoidal graphene samples with an edge crack [15]. (c) Sequential snapshots of crack propagation in the sinusoidal graphene [15]. (d) Crack paths in rubber sheets draped on curved substrates [147].**

The topology-induced toughening, or simply topological toughening, discussed in this part demonstrates that the essence of topological design of 2D materials lies with the intrinsic connection between out-of-plane deformation and topological defect distribution. This connection is highly nonlinear and often involves strong multi-physics coupling. While topological toughening could be a promising way to introduce controllable/designable toughening mechanisms into 2D materials to mitigate or overcome their intrinsic brittleness [148], much effort is needed in the future to explore the full potential of enhancing the mechanical and physical properties of graphene and other 2D materials through topological design.

## 3. Applications of topologically designed graphene

Beyond theoretical predictions of enhanced mechanical properties of graphene through topological design, topological defects have already been shown to play critical roles in a number of novel applications, such as chirality-specific single walled carbon nanotube growth, energy material engineering, multi-functional materials and interaction with biological systems. In this section, we will briefly review some of the advancements in this field. We emphasize that a rational design and fabrication of topological defects in graphene for specific applications have not yet been fully realized, and there will be tremendous opportunities to optimize the performance of novel devices and techniques based on topologically designed graphene once their fabrication techniques are matured.

*3.1 Topologically designed graphene flake[2] to guide the growth of single walled carbon nanotube (SWCNT)*

As an important one-dimensional nanomaterial, SWCNT has attracted a great deal of research interests due to its extraordinary physical properties and promising potential in various applications [149-151]. Many of these properties and applications are strongly dependent on the structure of the SWCNT, such as the diameter and the orientation angle of the hexagonal lattice relative to the tube axis, also known as chirality (*n*, *m*). For example, SWCNTs could be either metallic or semiconducting depending on the chirality [152]. In semiconducting SWCNTs, band gaps are inversely proportional to the diameter. Thus, fabrication of chirality-specific SWCNTs is of great importance to achieve the full technical potential of CNTs. Recently, it has been shown that a $C_{96}H_{54}$ precursor [153] could be transformed into a 3D curved nano-graphene flake with identical atomic structure as the end-cap of a (6, 6) nanotube through an intramolecular cyclodehydrogenation process on a Pt (111) surface. The (6, 6) caps could be used as seeds to grow defect-free SWCNTs with lengths up to a few hundred nanometers through a surface-catalyzed growth process (Fig 13a). The well-designed atomic structure of the 3D end-cap seeds enables the synthesis of SWCNTs with a specific chirality (6, 6) (more than 90%), instead of a mixture of uncontrolled structures. Through different synthesis routes, 3D topologically designed nano-graphene flakes with end-cap structures for (5,5), (9,0), (8,8), (10,10) and (12,12) SWCNTs have also been reported [154-156], further demonstrating the potential of topological design (Fig 13b-c). It remains a challenge to systematically design topological structures of the SWCNT end cap with different chiralities.

---

[2] Graphene flake here refers to nanoscale carbon molecules made of single layer $sp^2$-hybridized carbon atoms.

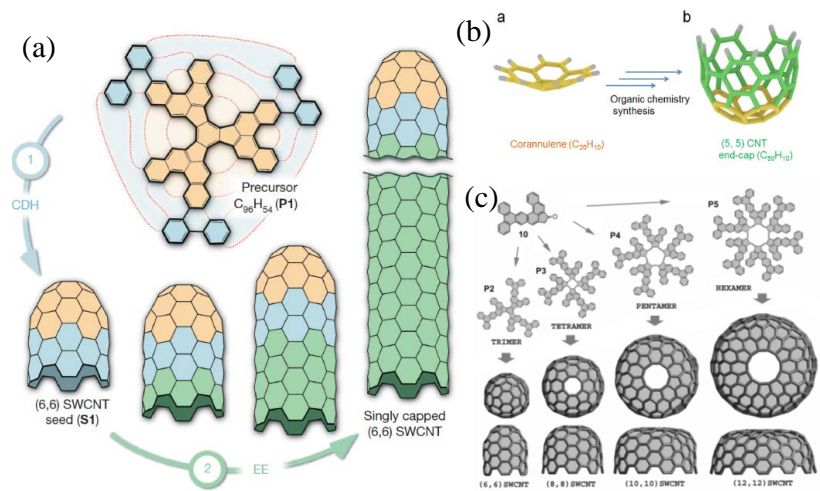

**Fig 13 Topologically designed graphene flake to guide the growth of chirality-specific SWCNTs. (a) Schematic illustration of a bottom-up synthesis of (6,6) SWCNTs from a designed end-cap [153]. (b) Atomic structure of the end-cap for (5,5) SWCNTs [155]. (c) Atomic structures of designed end-caps for (6,6), (8,8), (10,10) and (12,12) SWCNTs [154].**

*3.2 Topologically designed graphene for novel energy related applications*

For energy related applications, topologically designed graphene has been used to enhance performance of rechargeable lithium ion battery (LIB) and supercapacitor systems. One criterion for choosing anode materials for LIB is to achieve high specific charge capacity [157]. Topological defects have been predicted to be useful in improving the capacity in graphene-made electrode. First-principle calculations [158] based on density functional theory (DFT) revealed that the lithium adsorption on graphene could be enhanced by topological defects including divacancy (5-8-5 rings) and Stone-Wales defect (5-7-5-7 rings) due to the increased charge transfer between the adatom and defected sites in graphene. Later theoretical studies found that not only topological defects like 5-, 7- and 8- rings (Fig 14a-c), but curvatures (Fig 14d) of the graphene sheet could also enhance lithium adsorption resulting in better lithium storage property [159].

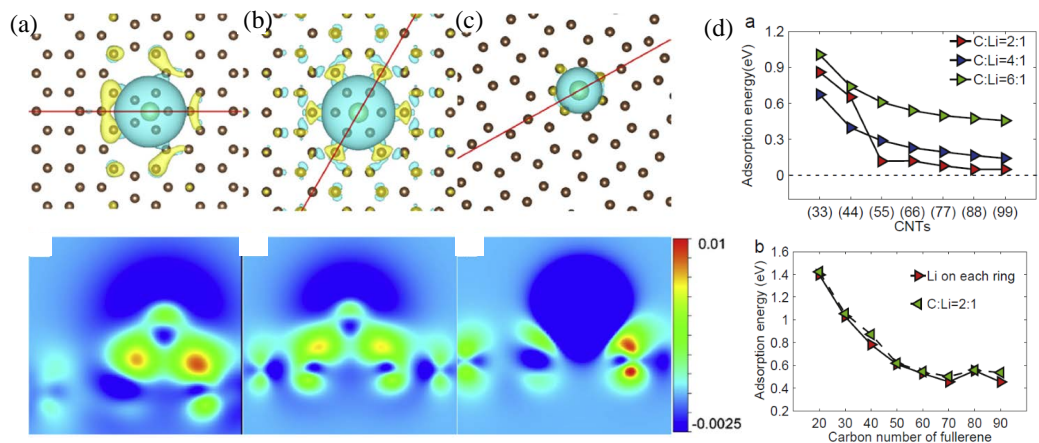

Fig 14 Lithium adsorption on graphene enhanced by topological defects and curvature [159]. (a)-(c) Top and side views of charge density of Li adsorbed in a pentagon ring (a), a hexagon ring (b) and a heptagon ring (c). Li on the defective rings (a) and (c) transfer more charge to C than on the pristine hexagon ring (b). (d) Adsorption energy of Li atoms adsorbed on CNTs and fullerene molecules is observed to increase with surface curvature.

Besides potential applications as an anode material, topologically designed graphene could also be integrated with other anode materials like silicon to optimize battery performance. Although silicon is known to possess the highest theoretical charge capacity [160], it suffers from chemo-mechanical degradations due to a large volume change (300%) during battery operation [161-163]. Fracture, loss of electrical contact and repeated chemical side reaction with the electrolyte can occur during lithiation and delithiation processes and a tremendous amount of research effort has been devoted to resolving these issues [164-170]. In a recent experimental study, Li et al. [171] designed 3D curved graphene cages to encapsulate micro-Si particles and achieved outstanding long time stability of the resulting anode. The 3D graphene cages possessing wavy profiles are mechanically strong and flexible; During battery operations, the encapsulated Si particles could undergo large deformation and even fracture without losing electrical contact because of the constraint of the graphene cages (Fig 15a-c). Additionally, it was shown that the solid electrolyte interphase (SEI) layer formed on a graphene cage remains intact during repeated lithiation/delithiation, resulting in stable cycling with 90% capacity retention after 100 cycles.

Topological defects and the resulting 3D curvatures have also been shown to contribute to the performance of electrochemical supercapacitors [172] made of graphene. DFT calculations [173] have predicted that topological defects such as 5-7-5-7 rings, 5-8-5 rings could substantially enhance the quantum capacitance of graphene by inducing quasi-localized states near the Fermi level, achieving a nearly 4-fold increase [174] in double-layer capacitance after combining with functionalization (Fig 15d).

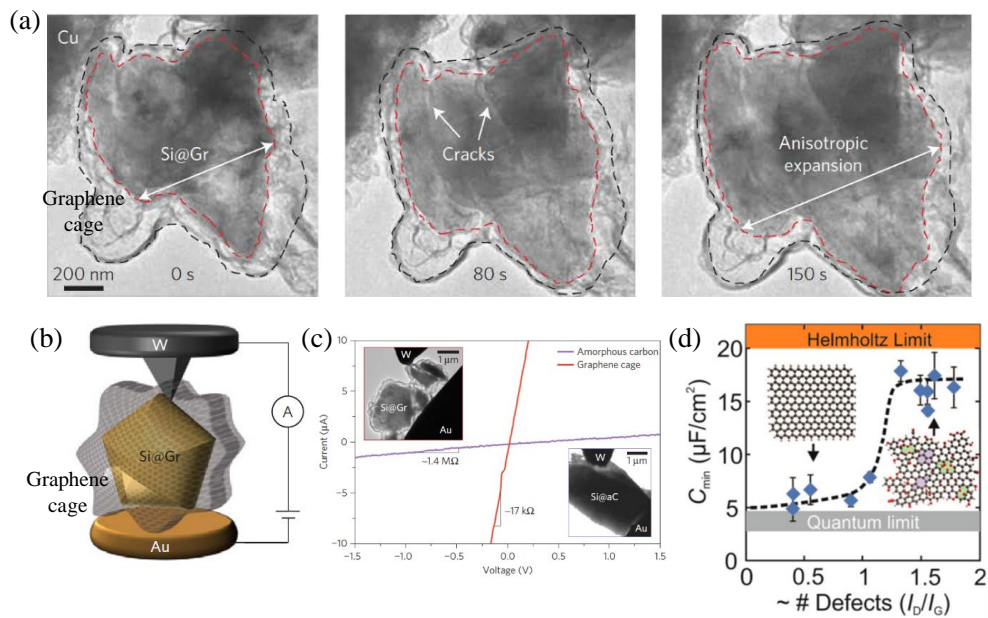

**Fig 15 Topologically designed graphene enhances the performance of Si anode and supercapacitors. (a) In situ TEM observation of deformation and fracture of a graphene caged Si particle during lithiation. The Si particle (outline in red) fractures abruptly and violently within the mechanically strong graphene cage (outlined in black) which remains intact throughout the process [171]. (b) Schematic diagram of a nanoscale electrochemical cell with a graphene caged Si particle [171]. (c) Current-voltage curve of graphene-encapsulated Si micro particle (SiMP) and amorphous-carbon-coated SiMP [171]. (d) Measured capacitance of graphene supercapacitor as a function of defect density [174].**

Note that in the studies discussed above, topological defects and 3D curvature serve as a new platform to couple mechanical deformation, chemical reaction and electronic structures of graphene in enhancing specific targeted properties via topological design.

### 3.3 Topologically designed graphene for multi-functional materials

Topological defects and 3D curvature have also shown promises in altering electrical transport behaviors [4,18,19,21], tuning thermal conductivity [16,22,175,176], generating mechanical-electrical coupling [56,57,177] and modifying chemical reactivity [17] in graphene systems. Yazyev and Louie [19] theoretically explored the potential of controlling electronic transport in graphene with GBs, finding two distinct transport behaviors, either high transparency or perfect reflection of charge carriers over remarkably large energy ranges, as shown in Fig 16a. In experiments, while Huang et al. [4] detected no measurable electrical resistance from GBs within instrument limits, Jauregui et al. [18] showed that GBs impede electrical transport and induce prominent weak localization, indicative of intervalley scattering in graphene. Tsen et al. [21] found that GBs with better inter-domain stitching lead to more uniform transport. These studies show that the effects of GBs on electrical properties of graphene are highly dependent on their atomic structures, suggesting possibilities to control electrical properties of graphene through grain boundary engineering.

On thermal properties of graphene, both theoretical studies and experiments have confirmed that the presence of topological defects could result in substantial reduction in thermal conductance [16,176,178]. This is mainly because phonons, as the dominant carriers of thermal energy in 2D materials, are scattered when they encounter topological defects, thereby limiting the phonon mean free path. This reduction in thermal conductance is closely related to the distribution of topological defects and can be characterized by the Kapitza resistance. Bagri et al. [175] studied thermal transport across several grain boundaries with different grain orientations using non-equilibrium molecular dynamics and found that the thermal conductance could be tuned in a certain range by the atomic structures of grain boundaries, as shown in Fig 16b. Serov et al. [16] found that the type and size of grain boundaries play important roles in determining the thermal conductance when grain sizes are smaller than a few hundred nanometers. Fthenakis et al. [176] showed that the thermal conductivity of graphene depends sensitively on whether the defects are isolated, form lines, or form extended arrangements in haeckelites. These studies point to the potential of tuning thermal properties of 2D materials through grain boundary engineering. Controlling heat conduction in materials is of great importance for thermal management of electronic devices as well as thermal energy recycling, which can be achieved through thermoelectric conversion that depends inversely on the thermal conductivity. Graphene has displayed larger Seebeck coefficient and higher overall power factor than other semiconductors and metals. By reducing the thermal conductivity of graphene while maintaining its electrical conductivity or increasing the ratio between electrical conductivity and thermal conductivity through defect engineering, the standard figure of merit for thermoelectricity (ZT) can be enhanced up to three times that of pristine graphene [179]. Ma et al. [22] experimentally measured the influence of grain size on the thermal and electrical transport behaviors in polycrystalline graphene, and further demonstrated the possibility of improving the thermoelectricity of 2D materials through grain size engineering.

Topological design can also play an important role in tailoring the piezoelectricity and flexoelectricity of graphene. Pristine graphene possesses no piezoelectric property due to its intrinsically centrosymmetric hexagonal structure. Piezoelectricity can be induced by breaking the structural symmetry, and doping is an effective way to create internal polarization [177]. Compared with piezoelectricity, flexoelectricity is a more universal phenomenon of dielectrics in which strain gradient can polarize the material and conversely, non-uniform electric fields can cause mechanical deformation. Based on DFT calculations of curved graphene surface, Kalinin and Meunier [57] proposed the concept of electronic flexoelectricity, in which the bending of single graphene layer results in a transfer of electron gas density across the basal plane and yields a curvature-dependent electrical dipole response. Later on, Kvashnin et al. [56] established the universality of the linear dependence of flexoelectric atomic dipole moments on local curvature in various carbon networks such as nanotubes, fullerenes, and nanocones. This field is still in its infancy with a lot of open questions. For example, since topological defects are intimately connected to the out-of-plane deformation and curvature of graphene, it will be interesting to investigate whether and how they induce flexoelectricity, and if they do, how to optimize such effect through topological design.

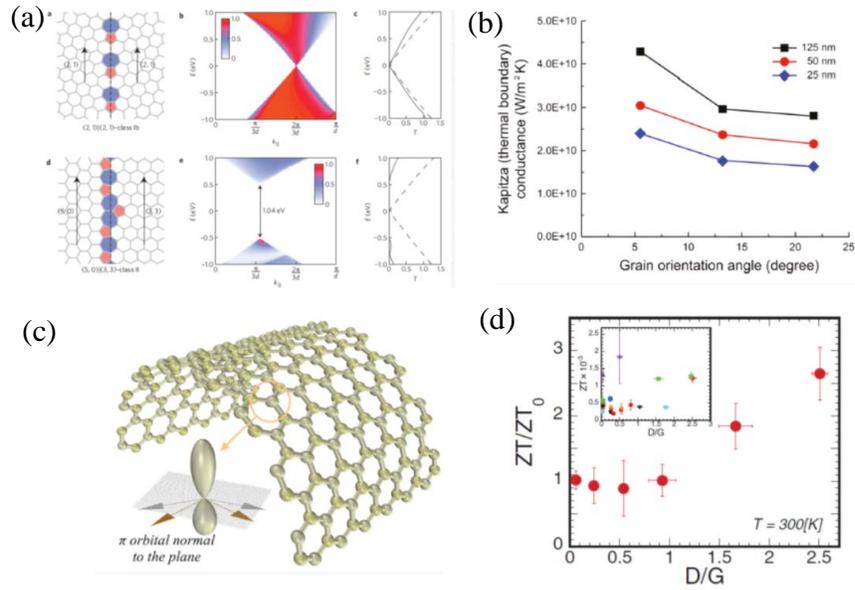

**Fig 16** Electrical, thermal, thermoelectrical, flexoelectrical properties of graphene with topological defects. (a) Two distinct electronic transport behaviors through GBs in graphene from first principles [19]. (b) Boundary conductance of GBs as a function of grain orientation angle [175]. (c) Curvature induced polarization upon bending [180]. (d) Normalized ZT values with respect to defect density [179].

The chemical reactivity of graphene has also been shown to be sensitive to the presence of topological defects and curvature. Through experimental study of curved nano-porous graphene, Ito et al. [17] have shown that highly curved graphene with a high density of topological defects can promote chemical doping contents, either electron donors or acceptors. Wu et al. [93] have shown that in situ generated aryl radicals were more likely to be found in regions with higher local curvature, demonstrating the selective effect of curvature on surface functionalization of graphene. Based on a molecular mechanics model, Pacheco Sanjuan et al. [181] found that the pyramidalization angle, which is directly proportional to the mean curvature of a curved 3D graphene, is important to determine the chemical reactivity (such as chemisorption of $H_2$) of the material. This theoretical study sheds light on the experimental observations and suggests a route to rationally design the chemical reactivity of graphene by controlling its mean curvature.

Potential use of topological design in tuning multi-functionalities of graphene has also been explored in bulk materials made of graphene. For example, Qin et al. [182] and Jung et al. [23] have studied the properties of 3D graphene with gyroid-shaped unit cells using MD simulations and 3D printed models. It was shown that the graphene gyroids could achieve high specific strength and low thermal conductivity due to the curvature and topological defects. Indeed, topological defects are geometrically necessary in 3D systems made of 2D materials, where topological design can be expected to play essential roles.

*3.4 Topologically designed graphene for biological applications*

Due to their remarkable multi-physical properties and large specific surface areas, graphene and its derivatives, such as CNTs, fullerene molecules and graphene oxide (GO), have been applied to various biological applications as diverse as biosensors [183,184], drug delivery [185,186] and biological imaging [187,188]. Experimental and theoretical studies have demonstrated that topological features, especially curvature, play important roles in determining the interaction of graphene with proteins [189-191], nucleic acids (such as DNA) and cell membranes, where weak non-covalent bonding is the dominant force of interaction.

Previous studies have found that proteins and DNA molecules interact with graphene mainly through $\pi$-$\pi$ stacking and dispersion interactions. The surface curvature of graphene has been proven to play important roles in modifying these non-covalently bonding interactions and tuning its adsorption capacity. For proteins interacting with graphene, Zuo et al. [189] demonstrated from MD simulations that the flat and flexible surface of graphene has better chance to form flat $\pi$-$\pi$ stacking with aromatic residues in protein villin headpiece (HP35) while the convex surfaces of SWCNT and $C_{60}$ interact with HP35 mainly through the dispersion interaction (Fig 17a-c). Beyond the flat and convex surfaces, Jana et al. [190] studied curvature dependence of polypeptide adsorption on flat, convex and concave graphene surfaces, and found that the concave surface shows the strongest absorptivity (Fig 17d). Gu et al. [191] studied the adsorption of bovine serum albumin (BSA), a model protein, on SWCNTs and graphene using MD simulations and fluorescence spectroscopy experiments, with results demonstrating that the adsorption capacity of the protein depends on surface curvature. Similar phenomena have also been reported for the interaction of graphene with DNA/RNA nucleobases and molecules. Gao et al. [192] showed that a DNA molecule could be spontaneously inserted into a SWCNT in aqueous solution using MD simulations (Fig 17e). Umadevi and Sastry [193] demonstrated that the binding energy of DNA/RNA nucleobases on the outer surface of a SWCNT increases with the radius of the SWCNT through quantum chemical calculations (Fig 17f). The curvature dependence of biomolecular adsorption could lead to novel devices to detect different molecules and even to remove harmful molecules in disease treatment.

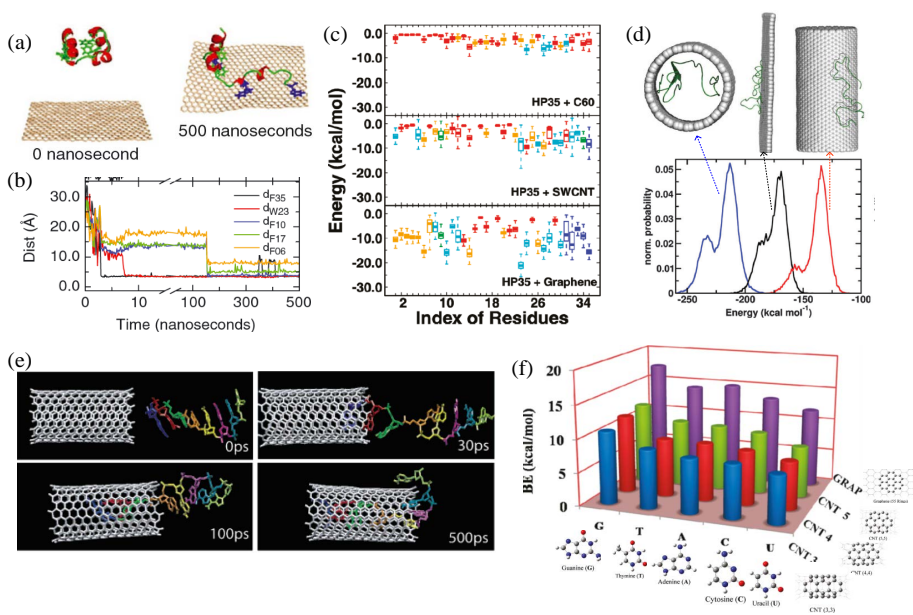

Fig 17 Curvature-dependent interactions between proteins, DNA molecules and graphene surfaces. (a) Representative snapshots of HP35 adsorbed on graphene surface [189]. The protein is shown in cartoons with red helix and green loop, and graphene is shown in orange. Aromatic residues that form π–π stacking are shown as blue stick, while the rest shown in green. (b) Distances between graphene and aromatic residues, including F35, W23, F10, F17 and F06 [189]. (c) Interaction energies between different residues of HP35 and graphene, (5,5)-SWCNT, and C60. Color of points indicates probability of a residue in contact with graphene: 0–20% (red), 20–40% (orange), 40–60% (green), 60–80% (cyan), and 80–100% (blue) [189]. (d) Normalized distribution of interaction energy of amphiphilic full-length amyloid beta peptide with concave (blue), flat (black), convex (red) graphene surfaces [190]. (e) Simulation snapshots of a DNA oligonucleotide interacting with a (10, 10) CNT in water solute environment [192]. (f) Binding energy of DNA/RNA nucleobase G, T, A, C, U with curved outer surface of CNTs and flat graphene [193].

Recent studies have demonstrated that graphene nanosheets could damage bacterial cell membrane by insertion/cutting as well as destructive extraction of lipid molecules [194-196]. This reveals a new toxic mechanism of graphene and opens the possibility of exploiting graphene as a novel antibacterial material [197,198]. The surface curvature of graphene is also found to affect lipid extraction via modifying the dispersive adhesion between graphene and lipid molecules. For example, using MD simulations and theoretical analysis, Luan et al. [199] have demonstrated that the lipid extraction could be understood as a wetting process and concave graphene surfaces possess the strongest extraction effect (Fig 18).

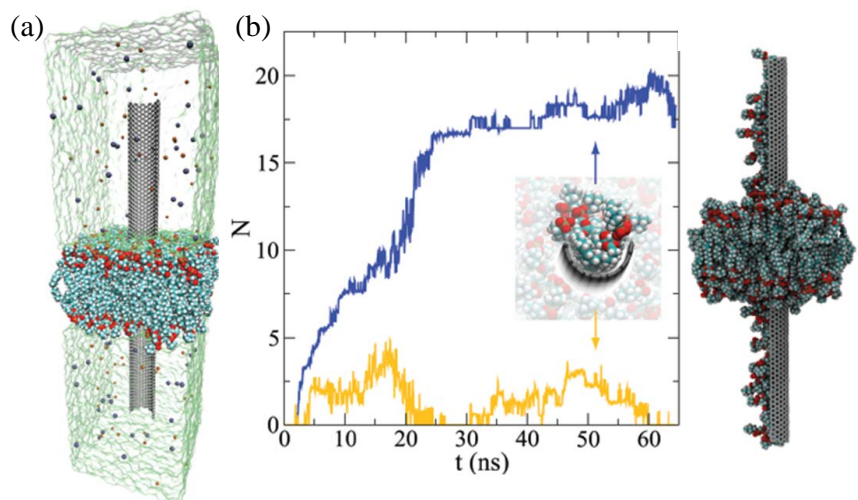

**Fig 18 Curvature dependent lipid extraction on graphene surfaces [199]. (a) MD simulation of curved graphene inserted into the bi-layer lipid membrane. (b) Time-dependent numbers of extracted lipids on a concave surface (blue) and a convex surface (orange) of graphene**

These studies clearly show a universal effect of surface curvature on the interactions of graphene with biomolecules, which is critical not only for applications of graphene in biotechnology but also in understanding the bio-safety of nanomaterials. Topological design may provide an effective means to control the surface curvature and morphology of graphene and further tune these interactions. Novel techniques and devices made of topological designed graphene could be expected to further unleash the potential in this direction.

## 4. Fabrication techniques of topologically designed graphene

While topologically designed graphene has been found promising in various applications, as discussed in Section 3, it is still very challenging and interesting to develop effective fabrication techniques to deliberately control the distributions of topological defect in graphene and other 2D materials. With the rapid development of capabilities in 2D materials synthesis, characterization and modification, several manipulation techniques and fabrication pathways have emerged as promising candidates for large scale applications of topologically designed graphene, such as CVD growth on curved templates, controlled irradiation and organic chemical synthesis.

First, controlled CVD growth continues to show promises for engineering topologically designed graphene at large scales. As a popular method to grow large scale graphene, currently CVD methods [4,5] can only produce samples with randomly distributed topological defects, including grain boundaries. However, with more controlled growth conditions, CVD methods could be further developed to generate desired patterns of topological defects. For example, pre-patterned growth seeds [81,200] may be used to tune the grain boundary density and patterns in polycrystalline graphene grown by CVD (Fig 19a). With appropriate substrate materials and growth conditions, graphene could grow on substrate of various geometrical and topological configurations, including porous metal foams [24,201], network of nanowires [202] and microparticles [191], 3D-printed scaffolds [203] and even zeolite crystal with nanoscale pores [204] (Fig 19a-f). To conform to the curvature of these surfaces, topological defects [24] are generated naturally and observed experimentally (Fig 19g). CVD growth of graphene on a curved substrate could be a promising way to fabricate topologically designed graphene at large scale. Similar attempts in CVD growth of TMDCs [205] have proven to be successful. For example, $WS_2$ growing on a cone-shape surface has been observed to yield grain boundaries [205].

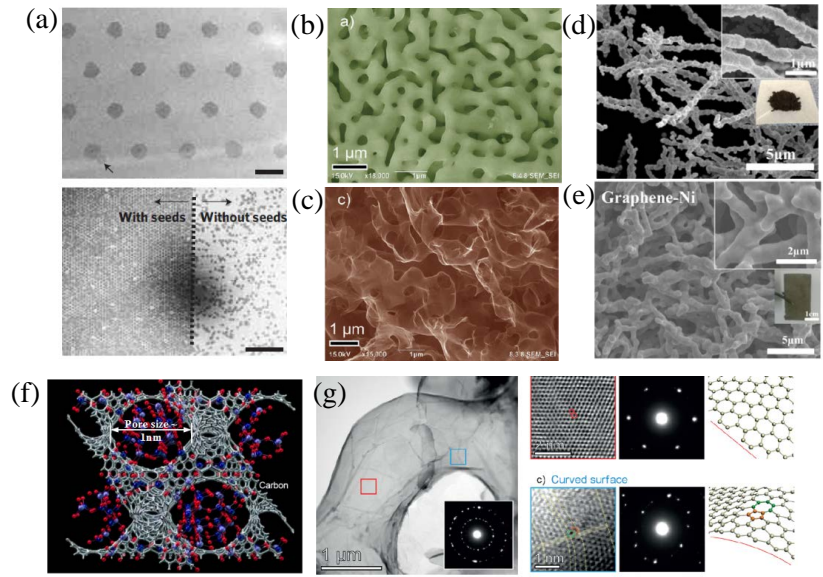

**Fig 19** Curved graphene grown on substrates with various geometries. (a) Seeded growth of graphene grains on a flat Cu foil [200]. (b)-(c) Nano-porous graphene (c) grown on the curved surface of a nano-porous Ni foam (b) [24]. (d)-(e) 3D graphene (e) grown on the surface of a Ni nanowire network (d) [202]. (f) Schematic of 3D graphene grown on the surface of zeolite crystal with nano pores [204]. (g) Topological defects and lattice bending in curved graphene grown on porous Ni foam surfaces [24].

Second, controlled irradiation or thermal excitation has been found to be able to create topological defects with desired type and locations at the atomic level. Early theoretical study [206] indicated that in the vicinity of vacancies, the energy barriers of the formation of various topological defects are sufficiently low in graphene that topological defects could be introduced via thermally activated reconstruction. Recent experiments have confirmed this prediction. For example, by adjusting the irradiation dosage and focus region of the electron beam, Robertson et al. [207] demonstrated that stable topological defects like dislocation pairs could be created within a controlled area with high spatial precision (~10*10 nm$^2$) (Fig 20a). Warner et al. [12] showed that with a slowly moving electron beam, a large number of dislocations could be created within a well-defined nanoscale area without creating holes in graphene (Fig 20b). Besides reorganizing the original carbon atoms via high-energy beams, it is also possible to add extra carbon atoms into graphene to create topological defects. Using a standard carbon coater, Lehtinen et al. [27] implanted extra carbon atoms into graphene samples to create dislocation dipoles, which can form atomic scale blisters with strong out-of-plane buckling (Fig 20c).

This seems to be a promising method to manipulate local configurations of topological defects, if higher precisions can be achieved in controlling detailed defect types.

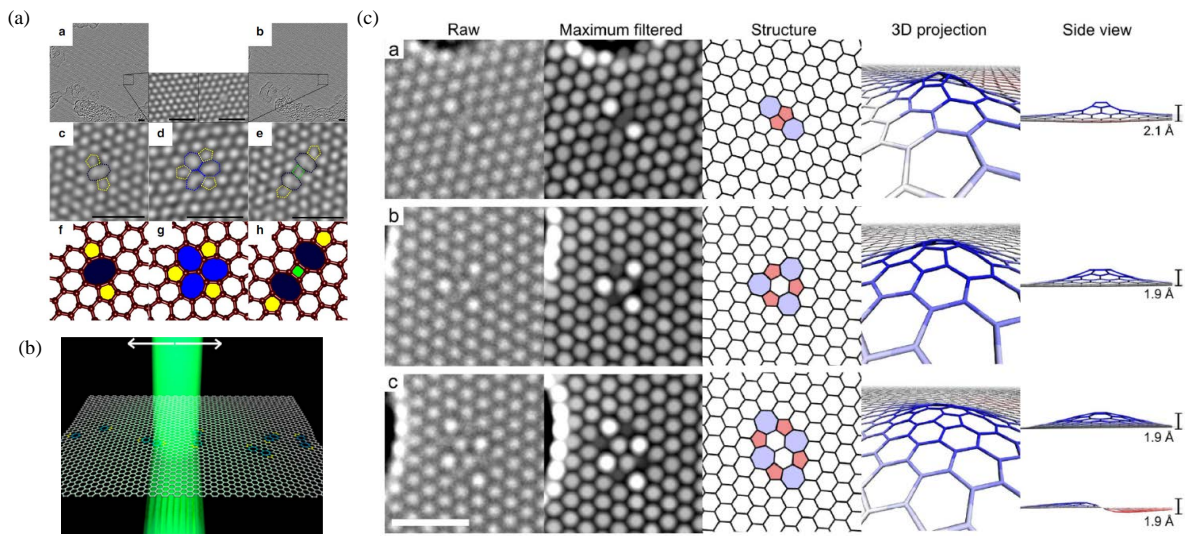

**Fig 20 Creating topological defects in graphene via controlled irradiation or thermal excitation. (a) Formation of topological defects in graphene after a 30s electron beam exposure [207]. (b) Schematic illustration of controlled creation of large dislocation numbers in graphene by scanning electron beam irradiation [12]. (c) Implantation of atomic scale blisters in graphene by depositing extra carbon atoms into single-layer graphene using a standard carbon coater [27].**

Third, at the molecular level, effective organic chemical synthesis pathways have been developed for different $sp^2$ carbons including curved graphene-like carbon with non-hexagon rings. For instance, chemical synthesis routes have been proposed and tested to create nanographene molecules with heptagon, pentagon and even octagon rings [26,208,209], whose equilibrium configurations are warped in 3D space (Fig 21a-c). Another study [153] has demonstrated that designed carbon molecules with both hexagons and non-hexagon rings are capable of folding into curved 3D configurations, like a spherical cap which could serve as a basis to grow carbon nanotubes with controlled chirality. At the same time, there has been a long-term trend of research on fabricating $sp^2$ carbon hybrids of fullerene molecules, carbon nanotubes and graphene. For example, seamless covalent bonding connection has been achieved during sequential growth of graphene layers and CNTs [25,204] (Fig 21d). Since the components in such hybrids show different curvatures, topological defects are naturally observed in the connection regions within these hybrids to accommodate the transition in curvature [25] (Fig 21e). The chemical synthesis routes and fabrication procedures could provide bottom-up pathways to engineer topological defects by combining growth and merging of small molecules.

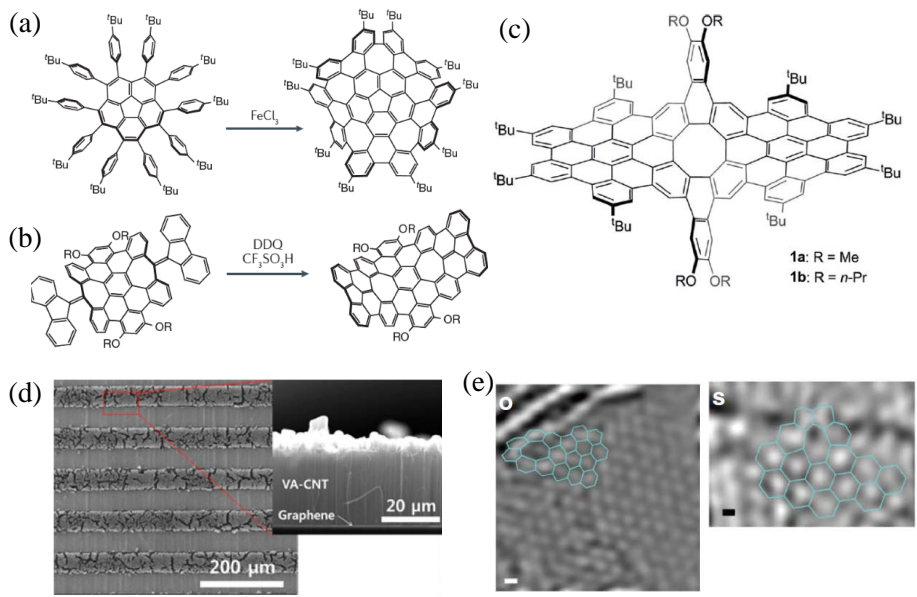

**Fig 21 Molecular level fabrications of nano-graphene flakes and graphene-CNT hybrids. (a)-(c) Examples of the syntheses of curved molecular nanographene flakes with 5-, 7- and 8-member rings [26,208,209]. (d) Selective growth of vertically aligned CNTs on patterned graphene [204]. (e) Topological defects at connections of seamless CNT-graphene hybrids [25].**

## 5. Outlook

In this chapter, we have reviewed some of the recent progresses in experimental and theoretical studies on graphene with topological defects and structures, summarizing some of the important and unique roles played by topological defects in atomically thin graphene structures. Based on these studies, one might propose topological design of 2D materials as an important research field to explore and take advantage of the full potential of graphene and other 2D materials. We have emphasized that topological defects play very important roles in determining the 3D curved geometry, residual stress field, as well as phonon and electron transport properties of graphene, which in principle could be designed for optimized mechanical, physical and chemical properties. The potential of topological design has been demonstrated through some of the preliminary successes in this field, such as tuning the strength of polycrystalline graphene, designing the shape of 3D curved graphene, enhancing fracture toughness of graphene, growing chirality-specific SWCNTs, engineering materials for energy related applications, designing multi-functional materials and tuning interactions with biological systems. In spite of the rapid developments in experiments and simulations, our understanding of the fundamental relations between topological defects and the mechanical, physical and chemical properties of 3D curved graphene and other 2D materials is still in its infancy. There remain many important open questions related to topological design of graphene that deserve further research efforts. Some of the open questions, opportunities and potential research topics/directions in this promising field are summarized below.

1. How can we systematically enhance the fracture toughness of graphene via topological design? In Section 2.3, we have reviewed some preliminary studies showing that various toughening mechanisms can be introduced in graphene via topological defects, including crack tip shielding by defects [140], stress reduction by defects-induced 3D geometry [15], crack branching, atomic-chain bridging [144] and nano-crack shielding [14]. These mechanisms lead to substantial enhancement in fracture toughness. Further improvement in toughness may be achieved by activating more effective toughening mechanisms or maximizing the synergistic effect of existing toughening mechanisms. Inspirations from biological materials and bulk engineering materials include crack trapping, shielding, deflection, crack bridging in nacre [210-212] and plasticity-induced energy dissipation in metals [213-215]. It will be interesting and challenging to see if it is possible introduce more toughening mechanisms (and even ductile deformation modes) into graphene via topological design. At the same time, theoretical models of topological toughening may need to touch on fracture mechanics of curved membrane and shell structures [216-218] to account for the intrinsic coupling between topological defects and out-of-plane geometry.

2. How does thermal fluctuation affect the topological design of graphene? Thermal fluctuation can significantly influence the mechanical and physical properties of graphene because of its very small bending stiffness (~ 1eV). At room temperature, it has been shown from self-consistent theory and MD simulations [61] that the stiffness of a micron scale graphene sheet can be enhanced by several orders of magnitude through thermal vibration, which seems to be supported by experiments on graphene kirigami [38]. An important general question is whether and how the 3D geometry of topologically designed graphene is influenced by thermal fluctuations. Besides directly modifying the effective properties of graphene, thermal fluctuations can also add time-varying perturbations to the morphology and mechanical properties (i.e., modulus and bending stiffness) of topologically designed graphene as they undergo random vibrations at finite temperature. These stochastic effects may be posed as an optimization problem with uncertainty.

3. How can one best tailor the multiphysical properties of graphene through topological design? By deliberately introducing topological defects, topological design can activate the effects of nanoscale straining and buckling on electronic band structure and chemical bonding in graphene. The topological defect-mediated coupling between electronic band structure, chemical reactivity, thermal conductivity, lattice distortion and 3D geometry raises interesting questions and challenges for understanding and manipulating multi-physical properties of graphene: (1) How does the topological design alter the electronic band structure [19], pseudomagnetic field [94] and surface plasmons [219,220] in graphene? (2) How do the curvature and residual stress introduced by topological defects interact with functional groups or corrosive agents? And what are the chemically stable topological designs in an ambient environment? (3) Can one further tune these properties of topologically designed graphene through controlled deformation or strain engineering [221,222]?

4. How can one design/achieve an optimal balance between various mechanical, physical and chemical properties in graphene/graphene devices through topological design? Numerous examples have shown that topological design of graphene could modify multiple properties simultaneously via the fully coupled mechanical, chemical and electrical interactions. Such coupling suggests that topological design of graphene can be a multi-objective optimization problem. For example, the out-of-plane geometry and curvature could generate electrical dipole moments through flexoelectricity [56,57], and the curvature-dependent polarization could be used in DNA sequencing [223,224]. At the same time, for large scale applications of such devices, it will be important to enhance the fracture toughness of graphene, which also depends on the out-of-plane geometry [15]. A design question might involve how to take the most advantage of flexoelectricity effect in DNA sequencing application while maintaining sufficient fracture toughness. It should be noticed that topological defects may diminish some properties while optimizing targeted ones. For example, defects can be used to enhance toughness of graphene [14,15], while they can also result in reduction of strength and/or modulus of graphene [8-11,54,131]. Therefore, another important question is how one can achieve optimally balanced properties with topological design.

5. Can topological design go beyond a single layer of graphene to an assembly of a large number of graphene structures, such as multi-layer graphene and 3D assemblies of interconnected graphene? For multi-layer graphene, strong differences exist between the intra- and inter-layer deformation, as the former is determined by the strong carbon-carbon covalent bonds, whereas the latter are governed by relatively weak forces (i.e. Van der Waals interaction). In addition, the inter-layer interaction is sensitive to atomic registry [225], surface geometry [54] and chemical bonding [226] between the layers. It could be expected that topological design could also affect the adhesion and friction between different layers. Fundamental and application questions could be raised in designing multi-layer graphene, including: How do the topological defects and curvature affect the inter-layer adhesion in multi-layer graphene systems? What are the friction mechanisms in multi-layer graphene with distributed topological defects? Can one tune or optimize the inter-layer adhesion and friction properties of multilayer graphene via topological design?

Graphene foam [24,203], CNT-graphene hybrids [25,204] and graphene gyroids [23,182] present examples of 3D interconnected graphene assemblies of graphene. Different from multi-layer graphene, 3D interconnected graphene consists of strong $sp^2$ carbon lattices spreading on 3D surfaces with complicated geometrical and topological configurations (e.g. gyroid surfaces), which occupies the whole 3D space. Curvature and topological defects play critical roles in maintaining the final self-equilibrium shapes and thus determine the effective bulk properties of the assembly. For example, by achieving a cork-like hierarchical structured 3D graphene monolith via freeze casting technique, Qiu et al. [226] demonstrated that the resulting 3D graphene foam with well-organized cellular structures could recover its shape and dimension after 90% compression. After 1,000 cycles of compression test, it only showed 7% reduction in dimension. This super-elastic behavior results from the well-organized structure of 3D graphene and is rarely achieved in poorly organized structures. Via MD simulations, Qin et al. [182] and Jung et al. [23] showed that Graphene gyroids can achieve excellent specific strength and unusually low thermal conductivity. These examples demonstrate the potential in exploiting geometrical/topological design of 3D interconnected graphene assemblies to further tune/optimize their properties and achieve novel applications. From the topological design point of view, there are plenty of challenges and opportunities beyond the monolayer topological design. For example, what is the best 3D surface of the interconnected graphene to optimize given properties, such as modulus-density ratio and thermal transfer? The necessary topological defects for forming these continuum surfaces will in turn alter the effective mechanical and physical properties and call for re-evaluation of the targeted functions. Therefore, another important question is how we can develop a fully coupled design methodology to integrate the global surface topology optimization and local topology dependent properties. Finally, the design schemes could also include the consideration of how to effectively synthesize such graphene structures with desired distribution of topological defects/3D geometry.

6. How can one efficiently and economically fabricate topologically designed graphene at different scales? In spite of the rapid developments in synthesizing various graphene structures with topological defects, as reviewed in Section 4, there are still a few key challenges in the application of the existing methods to achieve topological design at different length scales. For large-scale CVD methods, an important question is how to keep the precise substrate confinement of graphene during growth, considering that the high temperature in CVD growth [5] could decrease the adhesion between graphene and substrate and even destroy the designed geometrical features of the substrate [69]. To resolve problems like these, one needs a fundamental understanding of how curvature and topological defects affect the energetics of the nucleation, growth and coarsening of 2D materials. For methods focusing on the atomic or molecular levels, efficiency remains a trade-off against precision in large scale applications. Additionally, the scalability of the methods needs to be improved. These challenges themselves in turn present important research topics and could inspire further studies both in theoretical and experimental aspects.

7. Can one generalize the topological design concept to other 2D materials? Beyond graphene, topological defects have also been widely observed in other 2D materials, ranging from hexagonal boron nitride and black phosphorous to transition metal dichalcogenide monolayers like $MoS_2$, $WS_2$ and so on. Similar to graphene, the presence of topological defects in these 2D materials [72,227-230] not only induces out-of-plane deformation but also modifies physical properties compared with pristine samples. Moreover, the chemical bonding, dislocation core structures, mobility and interaction of topological defects can vary greatly for different 2D materials. For example, due to polarized bonding, grain boundaries in h-BN could carry net charges [90]. Because of the puckered lattice structure, dislocation cores in black phosphorous are predicted to have not only 5-7 but also 4-8, 5-8-7 and 5-8-8-7 rings [229]. Recently, an experimental study observed high-frequency dislocation emission in $MoS_2$ during crack propagation, suggesting possible plastic deformation activities in TMDCs [89]. These predictions and observations are broadening the horizon of topological design of 2D materials beyond graphene and suggest new opportunities and challenges for further study. For instance, (1) how could one modify the methods like PFC to solve the inverse design problem in topological design for 2D materials with multiple elements (e.g. h-BN) [231] or non-flat lattice (black phosphorous and TMDCs)? (2) What are the similarities and differences in the topological defect mediated coupling between the 3D curved geometry and multi-physical properties in different 2D materials? (3) What novel functionalities could be achieved through the topological design of a heterogeneous system combining different 2D materials?

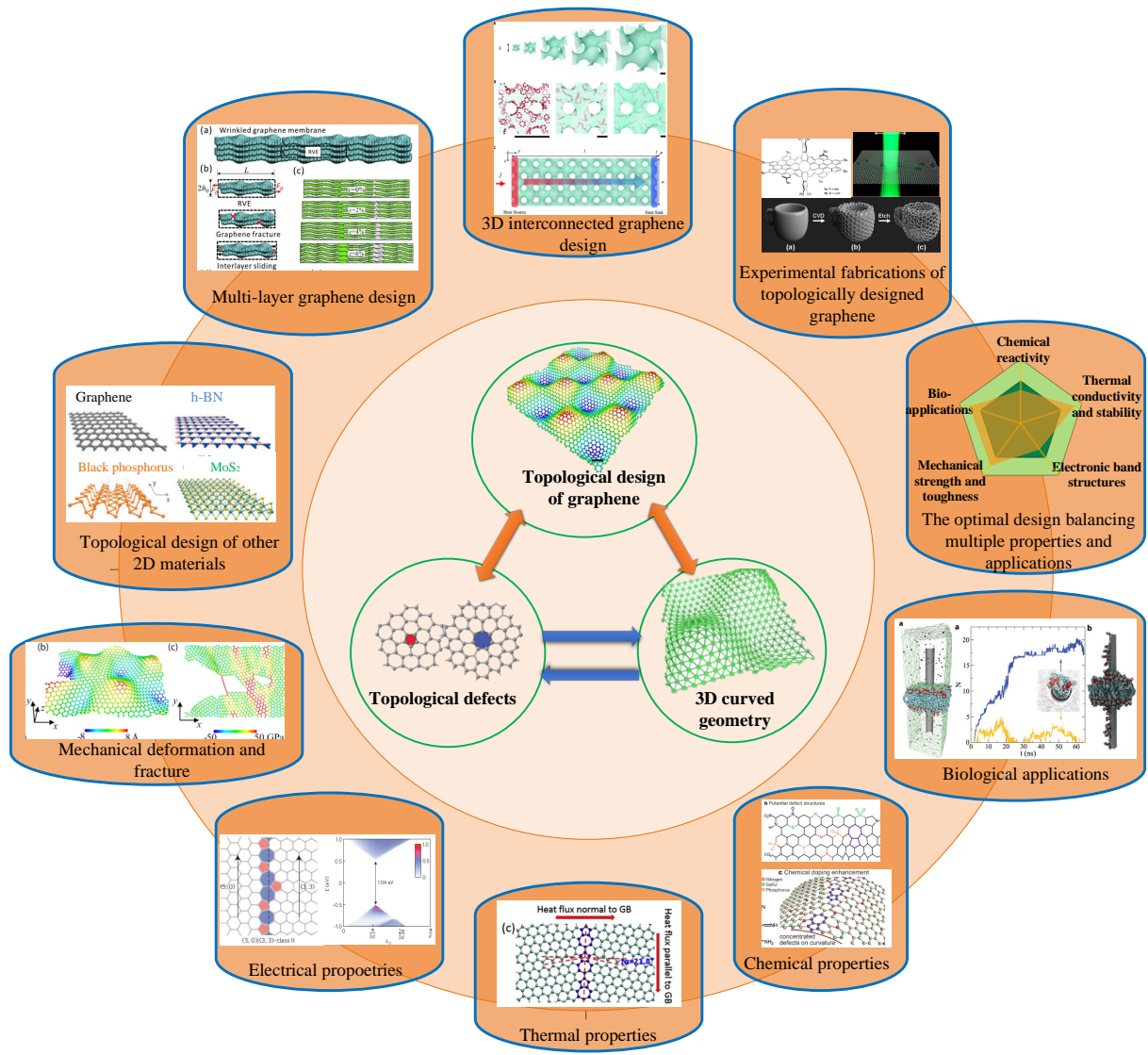

**Fig 22** Outlooks for topological design of graphene: morphology and curvature; strength and toughness; multifunction; nonlinear multi-physical coupling; multiscale fabrication; interconnected and multilayer graphene; extension to other 2D materials [12,15,17,23,26,28,54,69,199,232,233]

In summary, the concept of topological design of graphene discussed in this chapter focuses on the topological defect mediated coupling between 3D geometry, residual stress distribution, deformation, strength, fracture, and phonon and electron behaviors in graphene. In principle, a virtual simulation platform could be developed to optimize targeted properties of graphene by directly manipulating its topological structure. The rapid developments of experimental and fabrication techniques are paving ways to achieve topologically design of graphene from atom to device levels. This is inherently a highly interdisciplinary effort across multiple research communities including mechanics, nanotechnology, material science, physics, chemistry and even biology. There exist plenty of research opportunities in this emerging field, where fundamental studies could be expected to greatly expand our knowledge about 2D materials as well as open new avenues of application in novel devices based on 2D materials in the foreseeable future.

**References:**

[1] A. K. Geim and K. S. Novoselov, Nature materials **6**, 183 (2007).
[2] Y. Liu and B. I. Yakobson, Nano letters **10**, 2178 (2010).
[3] O. V. Yazyev and S. G. Louie, Physical Review B **81**, 195420 (2010).
[4] P. Y. Huang *et al.*, Nature **469**, 389 (2011).
[5] X. Li *et al.*, Science **324**, 1312 (2009).
[6] K. S. Kim *et al.*, nature **457**, 706 (2009).
[7] K. Kim, Z. Lee, W. Regan, C. Kisielowski, M. Crommie, and A. Zettl, ACS nano **5**, 2142 (2011).
[8] R. Grantab, V. B. Shenoy, and R. S. Ruoff, Science **330**, 946 (2010).
[9] Y. Wei, J. Wu, H. Yin, X. Shi, R. Yang, and M. Dresselhaus, Nature materials **11**, 759 (2012).
[10] H. I. Rasool, C. Ophus, W. S. Klug, A. Zettl, and J. K. Gimzewski, Nature communications **4**, 2811 (2013).
[11] A. Shekhawat and R. O. Ritchie, Nature communications **7**, 10546 (2016).
[12] J. H. Warner, Y. Fan, A. W. Robertson, K. He, E. Yoon, and G. D. Lee, Nano letters **13**, 4937 (2013).
[13] T. Zhang, X. Li, and H. Gao, Journal of the Mechanics and Physics of Solids **67**, 2 (2014).
[14] G. Jung, Z. Qin, and M. J. Buehler, Extreme Mechanics Letters **2**, 52 (2015).
[15] T. Zhang, X. Li, and H. Gao, Extreme Mechanics Letters **1**, 3 (2014).
[16] A. Y. Serov, Z.-Y. Ong, and E. Pop, Applied Physics Letters **102**, 033104 (2013).
[17] Y. Ito *et al.*, Advanced Materials **28**, 10644 (2016).
[18] L. A. Jauregui, H. Cao, W. Wu, Q. Yu, and Y. P. Chen, Solid State Communications **151**, 1100 (2011).
[19] O. V. Yazyev and S. G. Louie, Nature materials **9**, 806 (2010).
[20] H. Zhang, G. Lee, C. Gong, L. Colombo, and K. Cho, The Journal of Physical Chemistry C **118**, 2338 (2014).
[21] A. W. Tsen *et al.*, Science **336**, 1143 (2012).
[22] T. Ma *et al.*, Nature communications **8**, 14486 (2017).
[23] G. S. Jung, J. Yeo, Z. Tian, Z. Qin, and M. J. Buehler, Nanoscale **9**, 13477 (2017).
[24] Y. Ito *et al.*, Angewandte Chemie International Edition **53**, 4822 (2014).
[25] Y. Zhu *et al.*, Nature communications **3**, 1225 (2012).
[26] K. Y. Cheung, C. K. Chan, Z. Liu, and Q. Miao, Angewandte Chemie **129**, 9131 (2017).
[27] O. Lehtinen, N. Vats, G. Algara-Siller, P. Knyrim, and U. Kaiser, Nano letters **15**, 235 (2014).
[28] P. Kim, Nature materials **9**, 792 (2010).
[29] J. H. Warner, E. R. Margine, M. Mukai, A. W. Robertson, F. Giustino, and A. I. Kirkland, Science **337**, 209 (2012).
[30] G. R. Bhimanapati *et al.*, Acs Nano **9**, 11509 (2015).
[31] D. Akinwande *et al.*, Extreme Mechanics Letters **13**, 42 (2017).
[32] L. Song *et al.*, Nano letters **10**, 3209 (2010).
[33] K. Watanabe, T. Taniguchi, and H. Kanda, Nature materials **3**, 404 (2004).

[34]   Q. H. Wang, K. Kalantar-Zadeh, A. Kis, J. N. Coleman, and M. S. Strano, Nature nanotechnology **7**, 699 (2012).
[35]   M. Chhowalla, H. S. Shin, G. Eda, L.-J. Li, K. P. Loh, and H. Zhang, Nature chemistry **5**, 263 (2013).
[36]   P. Russo, A. Hu, and G. Compagnini, Nano-micro letters **5**, 260 (2013).
[37]   L. Jiang and Z. Fan, Nanoscale **6**, 1922 (2014).
[38]   M. K. Blees *et al.*, Nature **524**, 204 (2015).
[39]   Z. Qi, D. K. Campbell, and H. S. Park, Physical Review B **90**, 245437 (2014).
[40]   L. Xu, T. C. Shyu, and N. A. Kotov, ACS nano **11**, 7587 (2017).
[41]   T. C. Shyu, P. F. Damasceno, P. M. Dodd, A. Lamoureux, L. Xu, M. Shlian, M. Shtein, S. C. Glotzer, and N. A. Kotov, Nature materials **14**, 785 (2015).
[42]   S. Zhu and T. Li, ACS nano **8**, 2864 (2014).
[43]   V. B. Shenoy and D. H. Gracias, Mrs Bulletin **37**, 847 (2012).
[44]   K. Kumar, H. Van Swygenhoven, and S. Suresh, Acta Materialia **51**, 5743 (2003).
[45]   K. Lu, L. Lu, and S. Suresh, Science **324**, 349 (2009).
[46]   Y. Tian *et al.*, Nature **493**, 385 (2013).
[47]   Q. Huang *et al.*, Nature **510**, 250 (2014).
[48]   Q. Pan, H. Zhou, Q. Lu, H. Gao, and L. Lu, Nature **551**, 214 (2017).
[49]   O. Lehtinen, S. Kurasch, A. Krasheninnikov, and U. Kaiser, Nature communications **4**, 2098 (2013).
[50]   C. Lee, X. Wei, J. W. Kysar, and J. Hone, science **321**, 385 (2008).
[51]   Q. Lu, M. Arroyo, and R. Huang, Journal of Physics D: Applied Physics **42**, 102002 (2009).
[52]   Y. Wei, B. Wang, J. Wu, R. Yang, and M. L. Dunn, Nano letters **13**, 26 (2012).
[53]   Z. Song, V. I. Artyukhov, J. Wu, B. I. Yakobson, and Z. Xu, ACS nano **9**, 401 (2014).
[54]   H. Qin, Y. Sun, J. Z. Liu, and Y. Liu, Carbon **108**, 204 (2016).
[55]   A. Cortijo and M. A. Vozmediano, Nuclear Physics B **763**, 293 (2007).
[56]   A. G. Kvashnin, P. B. Sorokin, and B. I. Yakobson, The journal of physical chemistry letters **6**, 2740 (2015).
[57]   S. V. Kalinin and V. Meunier, Physical Review B **77**, 033403 (2008).
[58]   Z. Ni, H. Wang, J. Kasim, H. Fan, T. Yu, Y. Wu, Y. Feng, and Z. Shen, Nano letters **7**, 2758 (2007).
[59]   J. C. Meyer, A. K. Geim, M. I. Katsnelson, K. S. Novoselov, T. J. Booth, and S. Roth, Nature **446**, 60 (2007).
[60]   A. Fasolino, J. Los, and M. I. Katsnelson, Nature materials **6**, 858 (2007).
[61]   D. Wan, D. R. Nelson, and M. J. Bowick, Physical Review B **96**, 014106 (2017).
[62]   D. Yllanes, S. S. Bhabesh, D. R. Nelson, and M. J. Bowick, Nature Communications **8**, 1381 (2017).
[63]   F. Ahmadpoor, P. Wang, R. Huang, and P. Sharma, Journal of the Mechanics and Physics of Solids **107**, 294 (2017).
[64]   A. Košmrlj and D. R. Nelson, Physical Review E **89**, 022126 (2014).
[65]   Y. L. Zhong, Z. Tian, G. P. Simon, and D. Li, Materials Today **18**, 73 (2015).
[66]   Y. Chen, X. L. Gong, and J. G. Gai, Advanced Science **3** (2016).
[67]   V. P. Pham, H.-S. Jang, D. Whang, and J.-Y. Choi, Chemical Society Reviews **46**, 6276 (2017).
[68]   G. López-Polín, M. Ortega, J. Vilhena, I. Alda, J. Gomez-Herrero, P. A. Serena, C. Gomez-Navarro, and R. Pérez, Carbon **116**, 670 (2017).
[69]   P. M. Wilson, G. N. Mbah, T. G. Smith, D. Schmidt, R. Y. Lai, T. Hofmann, and A. Sinitskii, Journal of Materials Chemistry C **2**, 1879 (2014).
[70]   J. Červenka, M. Katsnelson, and C. Flipse, Nature Physics **5**, 840 (2009).
[71]   J. Lahiri, Y. Lin, P. Bozkurt, I. I. Oleynik, and M. Batzill, Nature nanotechnology **5**, 326 (2010).
[72]   A. L. Gibb, N. Alem, J.-H. Chen, K. J. Erickson, J. Ciston, A. Gautam, M. Linck, and A. Zettl, Journal of the American Chemical Society **135**, 6758 (2013).

[73] A. M. Van Der Zande *et al.*, Nature materials **12**, 554 (2013).
[74] Z. Zhang, Y. Yang, F. Xu, L. Wang, and B. I. Yakobson, Advanced Functional Materials **25**, 367 (2015).
[75] J. Wu and Y. Wei, Journal of the Mechanics and Physics of Solids **61**, 1421 (2013).
[76] N. Khosravian, M. K. Samani, G. C. Loh, G. C. K. Chen, D. Baillargeat, and B. K. Tay, Computational Materials Science **79**, 132 (2013).
[77] Z. D. Sha, S. S. Quek, Q. X. Pei, Z. S. Liu, T. J. Wang, V. B. Shenoy, and Y. W. Zhang, Scientific Reports **4**, 5991 (2014).
[78] Z. Song, V. I. Artyukhov, B. I. Yakobson, and Z. Xu, Nano letters **13**, 1829 (2013).
[79] A. T. Murdock *et al.*, ACS Nano **7**, 1351 (2013).
[80] X. Song *et al.*, Nano Letters **16**, 6109 (2016).
[81] D. Geng *et al.*, Proceedings of the National Academy of Sciences **109**, 7992 (2012).
[82] H. Shu, X. Chen, X. Tao, and F. Ding, ACS Nano **6**, 3243 (2012).
[83] K. Elder, M. Katakowski, M. Haataja, and M. Grant, Physical review letters **88**, 245701 (2002).
[84] J. Li, B. Ni, T. Zhang, and H. Gao, Journal of the Mechanics and Physics of Solids (2017).
[85] W. Guo *et al.*, ACS Nano **9**, 5792 (2015).
[86] J. Kotakoski and J. C. Meyer, Physical Review B **85**, 195447 (2012).
[87] Z. Sha, Q. Wan, Q. Pei, S. Quek, Z. Liu, Y. Zhang, and V. Shenoy, Scientific reports **4**, 7437 (2014).
[88] Z. Yang, Y. Huang, F. Ma, Y. Sun, K. Xu, and P. K. Chu, Materials Science and Engineering: B **198**, 95 (2015).
[89] T. H. Ly, J. Zhao, M. O. Cichocka, L.-J. Li, and Y. H. Lee, Nature communications **8**, 14116 (2017).
[90] Y. Liu, X. Zou, and B. I. Yakobson, ACS nano **6**, 7053 (2012).
[91] X. Zou, Y. Liu, and B. I. Yakobson, Nano letters **13**, 253 (2012).
[92] J. S. Choi *et al.*, Science **333**, 607 (2011).
[93] Q. Wu *et al.*, Chemical Communications **49**, 677 (2012).
[94] N. Levy, S. Burke, K. Meaker, M. Panlasigui, A. Zettl, F. Guinea, A. C. Neto, and M. Crommie, Science **329**, 544 (2010).
[95] V. M. Pereira, A. C. Neto, H. Liang, and L. Mahadevan, Physical review letters **105**, 156603 (2010).
[96] N. N. Klimov, S. Jung, S. Zhu, T. Li, C. A. Wright, S. D. Solares, D. B. Newell, N. B. Zhitenev, and J. A. Stroscio, Science **336**, 1557 (2012).
[97] P. Zhang *et al.*, Nature communications **5**, 3782 (2014).
[98] S. Chen and D. Chrzan, Physical Review B **84**, 214103 (2011).
[99] Y. Shibuta and S. Maruyama, Chemical physics letters **382**, 381 (2003).
[100] N. Piper, Y. Fu, J. Tao, X. Yang, and A. To, Chemical Physics Letters **502**, 231 (2011).
[101] E. Biyikli, J. Liu, X. Yang, and A. C. To, RSC Advances **3**, 1359 (2013).
[102] C. Chuang, Y.-C. Fan, and B.-Y. Jin, Journal of chemical information and modeling **49**, 361 (2009).
[103] V. Varshney, V. Unnikrishnan, J. Lee, and A. K. Roy, Nanoscale **10**, 403 (2018).
[104] T. C. Petersen, I. K. Snook, I. Yarovsky, and D. G. McCulloch, Physical Review B **72**, 125417 (2005).
[105] A. Hexemer, V. Vitelli, E. J. Kramer, and G. H. Fredrickson, Physical Review E **76**, 051604 (2007).
[106] L. Mitchell and A. Head, Journal of the Mechanics and Physics of Solids **9**, 131 (1961).
[107] H. Seung and D. R. Nelson, Physical Review A **38**, 1005 (1988).
[108] L. M. Zubov, *Nonlinear theory of dislocations and disclinations in elastic bodies* (Springer Science & Business Media, 1997), Vol. 47.
[109] L. Zubov, in *Doklady Physics* (Springer, 2007), pp. 67.
[110] L. Zubov, Journal of Applied Mathematics and Mechanics **74**, 663 (2010).

[111]  J. Dervaux and M. B. Amar, Physical review letters **101**, 068101 (2008).
[112]  H. Liang and L. Mahadevan, Proceedings of the National Academy of Sciences **106**, 22049 (2009).
[113]  S. J. Stuart, A. B. Tutein, and J. A. Harrison, The Journal of chemical physics **112**, 6472 (2000).
[114]  M. Diab, T. Zhang, R. Zhao, H. Gao, and K.-S. Kim, in *Proc. R. Soc. A* (The Royal Society, 2013), p. 20120753.
[115]  H. Emmerich, H. Löwen, R. Wittkowski, T. Gruhn, G. I. Tóth, G. Tegze, and L. Gránásy, Advances in Physics **61**, 665 (2012).
[116]  P. Galenko, H. Gomez, N. Kropotin, and K. Elder, Physical Review E **88**, 013310 (2013).
[117]  H. Qin, Y. Sun, J. Z. Liu, M. Li, and Y. Liu, Nanoscale **9**, 4135 (2017).
[118]  T. L. Anderson, *Fracture mechanics: fundamentals and applications* (CRC press, 2017).
[119]  G. Eda, G. Fanchini, and M. Chhowalla, Nature nanotechnology **3**, 270 (2008).
[120]  A. Reina, X. Jia, J. Ho, D. Nezich, H. Son, V. Bulovic, M. S. Dresselhaus, and J. Kong, Nano letters **9**, 30 (2008).
[121]  J. W. Suk, A. Kitt, C. W. Magnuson, Y. Hao, S. Ahmed, J. An, A. K. Swan, B. B. Goldberg, and R. S. Ruoff, ACS nano **5**, 6916 (2011).
[122]  J. Kang, D. Shin, S. Bae, and B. H. Hong, Nanoscale **4**, 5527 (2012).
[123]  L. Gao, G.-X. Ni, Y. Liu, B. Liu, A. H. C. Neto, and K. P. Loh, Nature **505**, 190 (2014).
[124]  M. Hofmann, Y.-P. Hsieh, A. L. Hsu, and J. Kong, Nanoscale **6**, 289 (2014).
[125]  K. Celebi, J. Buchheim, R. M. Wyss, A. Droudian, P. Gasser, I. Shorubalko, J.-I. Kye, C. Lee, and H. G. Park, Science **344**, 289 (2014).
[126]  Q. M. Ramasse, R. Zan, U. Bangert, D. W. Boukhvalov, Y.-W. Son, and K. S. Novoselov, ACS nano **6**, 4063 (2012).
[127]  Y. Hwangbo *et al.*, Scientific reports **4**, 4439 (2014).
[128]  R. Khare, S. L. Mielke, J. T. Paci, S. Zhang, R. Ballarini, G. C. Schatz, and T. Belytschko, Physical Review B **75**, 075412 (2007).
[129]  S. S. Terdalkar, S. Huang, H. Yuan, J. J. Rencis, T. Zhu, and S. Zhang, Chemical Physics Letters **494**, 218 (2010).
[130]  D. Cohen-Tanugi and J. C. Grossman, Nano letters **14**, 6171 (2014).
[131]  T. Zhang, X. Li, S. Kadkhodaei, and H. Gao, Nano letters **12**, 4605 (2012).
[132]  D. Sen, K. S. Novoselov, P. M. Reis, and M. J. Buehler, Small **6**, 1108 (2010).
[133]  M. J. Moura and M. Marder, Physical Review E **88**, 032405 (2013).
[134]  X. Huang, H. Yang, A. C. van Duin, K. J. Hsia, and S. Zhang, Physical Review B **85**, 195453 (2012).
[135]  S. Zhao and J. Xue, Journal of Physics D: Applied Physics **46**, 135303 (2013).
[136]  L. Lu, X. Chen, X. Huang, and K. Lu, Science **323**, 607 (2009).
[137]  X. Li, Y. Wei, L. Lu, K. Lu, and H. Gao, Nature **464**, 877 (2010).
[138]  W.-J. Moon, T. Ito, S. Uchimura, and H. Saka, Materials Science and Engineering: A **387**, 837 (2004).
[139]  H. Saka, Philosophical magazine letters **80**, 461 (2000).
[140]  F. Meng, C. Chen, and J. Song, The journal of physical chemistry letters **6**, 4038 (2015).
[141]  A. Verma and A. Parashar, Physical Chemistry Chemical Physics **19**, 16023 (2017).
[142]  G. Rajasekaran and A. Parashar, RSC Advances **6**, 26361 (2016).
[143]  S. Wang, B. Yang, J. Yuan, Y. Si, and H. Chen, Scientific reports **5**, 14957 (2015).
[144]  G. Rajasekaran and A. Parashar, Diamond and Related Materials **74**, 90 (2017).
[145]  J. Han, D. Sohn, W. Woo, and D.-K. Kim, Computational Materials Science **129**, 323 (2017).
[146]  Y. Wang and Z. Liu, Modelling and Simulation in Materials Science and Engineering **24**, 085002 (2016).
[147]  N. P. Mitchell, V. Koning, V. Vitelli, and W. T. Irvine, Nature materials **16**, 89 (2017).
[148]  T. Zhang and H. Gao, Journal of Applied Mechanics **82**, 051001 (2015).

[149] A. Jorio, G. Dresselhaus, and M. S. Dresselhaus, *Carbon Nanotubes: Advanced Topics in the Synthesis, Structure, Properties and Applications* (Springer Berlin Heidelberg, 2007).
[150] D. Jariwala, V. K. Sangwan, L. J. Lauhon, T. J. Marks, and M. C. Hersam, Chemical Society Reviews **42**, 2824 (2013).
[151] J. Wang, Electroanalysis **17**, 7 (2005).
[152] T. W. Odom, J.-L. Huang, P. Kim, and C. M. Lieber, (ACS Publications, 2000).
[153] J. R. Sanchez-Valencia, T. Dienel, O. Gröning, I. Shorubalko, A. Mueller, M. Jansen, K. Amsharov, P. Ruffieux, and R. Fasel, Nature **512**, 61 (2014).
[154] A. Mueller and K. Y. Amsharov, European Journal of Organic Chemistry **2015**, 3053 (2015).
[155] B. Liu *et al.*, Nano letters **15**, 586 (2014).
[156] N. Abdurakhmanova, A. Mueller, S. Stepanow, S. Rauschenbach, M. Jansen, K. Kern, and K. Y. Amsharov, Carbon **84**, 444 (2015).
[157] J.-M. Tarascon and M. Armand, in *Materials For Sustainable Energy: A Collection of Peer-Reviewed Research and Review Articles from Nature Publishing Group* (World Scientific, 2011), pp. 171.
[158] D. Datta, J. Li, N. Koratkar, and V. B. Shenoy, Carbon **80**, 305 (2014).
[159] Z. Pang, X. Shi, Y. Wei, and D. Fang, Carbon **107**, 557 (2016).
[160] D. Lin, Y. Liu, and Y. Cui, Nature nanotechnology **12**, 194 (2017).
[161] L. Beaulieu, K. Eberman, R. Turner, L. Krause, and J. Dahn, Electrochemical and Solid-State Letters **4**, A137 (2001).
[162] M. Obrovac and L. Christensen, Electrochemical and Solid-State Letters **7**, A93 (2004).
[163] M. Obrovac, L. Christensen, D. B. Le, and J. R. Dahn, Journal of The Electrochemical Society **154**, A849 (2007).
[164] C. K. Chan, H. Peng, G. Liu, K. McIlwrath, X. F. Zhang, R. A. Huggins, and Y. Cui, Nature nanotechnology **3**, 31 (2008).
[165] L.-F. Cui, R. Ruffo, C. K. Chan, H. Peng, and Y. Cui, Nano letters **9**, 491 (2008).
[166] S. Zhou, X. Liu, and D. Wang, Nano letters **10**, 860 (2010).
[167] M.-H. Park, M. G. Kim, J. Joo, K. Kim, J. Kim, S. Ahn, Y. Cui, and J. Cho, Nano letters **9**, 3844 (2009).
[168] J. Xiao, W. Xu, D. Wang, D. Choi, W. Wang, X. Li, G. L. Graff, J. Liu, and J.-G. Zhang, Journal of The Electrochemical Society **157**, A1047 (2010).
[169] A. Magasinski, P. Dixon, B. Hertzberg, A. Kvit, J. Ayala, and G. Yushin, Nature materials **9**, 353 (2010).
[170] N. Liu, H. Wu, M. T. McDowell, Y. Yao, C. Wang, and Y. Cui, Nano letters **12**, 3315 (2012).
[171] Y. Li, K. Yan, H.-W. Lee, Z. Lu, N. Liu, and Y. Cui, Nature Energy **1**, 15029 (2016).
[172] C. Liu, Z. Yu, D. Neff, A. Zhamu, and B. Z. Jang, Nano letters **10**, 4863 (2010).
[173] L.-J. Zhou, Z. Hou, and L.-M. Wu, The Journal of Physical Chemistry C **116**, 21780 (2012).
[174] M. A. Pope and I. A. Aksay, The Journal of Physical Chemistry C **119**, 20369 (2015).
[175] A. Bagri, S.-P. Kim, R. S. Ruoff, and V. B. Shenoy, Nano letters **11**, 3917 (2011).
[176] Z. G. Fthenakis, Z. Zhu, and D. Tománek, Physical Review B **89**, 125421 (2014).
[177] M. T. Ong and E. J. Reed, ACS nano **6**, 1387 (2012).
[178] P. Yasaei *et al.*, Nano Letters **15**, 4532 (2015).
[179] A. Yuki, I. Yuki, T. Kuniharu, A. Seiji, and A. Takayuki, 2D Materials **4**, 025019 (2017).
[180] F. Ahmadpoor and P. Sharma, Nanoscale **7**, 16555 (2015).
[181] A. A. Pacheco Sanjuan, M. Mehboudi, E. O. Harriss, H. Terrones, and S. Barraza-Lopez, ACS nano **8**, 1136 (2014).
[182] Z. Qin, G. S. Jung, M. J. Kang, and M. J. Buehler, Science advances **3**, e1601536 (2017).
[183] Y. Shao, J. Wang, H. Wu, J. Liu, I. A. Aksay, and Y. Lin, Electroanalysis **22**, 1027 (2010).
[184] J. Liu, Z. Liu, C. J. Barrow, and W. Yang, Analytica chimica acta **859**, 1 (2015).
[185] S. Goenka, V. Sant, and S. Sant, Journal of Controlled Release **173**, 75 (2014).
[186] K. Yang, L. Feng, and Z. Liu, Expert opinion on drug delivery **12**, 601 (2015).
[187] G. Hong, S. Diao, A. L. Antaris, and H. Dai, Chemical reviews **115**, 10816 (2015).

[188] J. M. Yoo, J. H. Kang, and B. H. Hong, Chemical Society Reviews **44**, 4835 (2015).
[189] G. Zuo, X. Zhou, Q. Huang, H. Fang, and R. Zhou, The Journal of Physical Chemistry C **115**, 23323 (2011).
[190] A. K. Jana, M. K. Tiwari, K. Vanka, and N. Sengupta, Physical Chemistry Chemical Physics **18**, 5910 (2016).
[191] Z. Gu, Z. Yang, Y. Chong, C. Ge, J. K. Weber, D. R. Bell, and R. Zhou, Scientific reports **5**, 10886 (2015).
[192] H. Gao, Y. Kong, D. Cui, and C. S. Ozkan, Nano Letters **3**, 471 (2003).
[193] D. Umadevi and G. N. Sastry, The Journal of Physical Chemistry Letters **2**, 1572 (2011).
[194] Y. Tu *et al.*, Nature nanotechnology **8**, nnano. 2013.125 (2013).
[195] S. Liu, M. Hu, T. H. Zeng, R. Wu, R. Jiang, J. Wei, L. Wang, J. Kong, and Y. Chen, Langmuir **28**, 12364 (2012).
[196] Y. Li, H. Yuan, A. von dem Bussche, M. Creighton, R. H. Hurt, A. B. Kane, and H. Gao, Proceedings of the National Academy of Sciences **110**, 12295 (2013).
[197] W. Hu, C. Peng, W. Luo, M. Lv, X. Li, D. Li, Q. Huang, and C. Fan, ACS nano **4**, 4317 (2010).
[198] S. Liu, T. H. Zeng, M. Hofmann, E. Burcombe, J. Wei, R. Jiang, J. Kong, and Y. Chen, ACS nano **5**, 6971 (2011).
[199] B. Luan, T. Huynh, and R. Zhou, Nanoscale **8**, 5750 (2016).
[200] Q. Yu *et al.*, Nature materials **10**, 443 (2011).
[201] Z. Chen, W. Ren, L. Gao, B. Liu, S. Pei, and H.-M. Cheng, Nature materials **10**, 424 (2011).
[202] B. H. Min, D. W. Kim, K. H. Kim, H. O. Choi, S. W. Jang, and H.-T. Jung, Carbon **80**, 446 (2014).
[203] Z. Yang, C. Yan, J. Liu, S. Chabi, Y. Xia, and Y. Zhu, RSC Advances **5**, 29397 (2015).
[204] K. Kim *et al.*, Nature **535**, 131 (2016).
[205] H. Yu, N. Gupta, Z. Hu, K. Wang, B. R. Srijanto, K. Xiao, D. B. Geohegan, and B. I. Yakobson, (2017).
[206] M. T. Lusk and L. D. Carr, Physical review letters **100**, 175503 (2008).
[207] A. W. Robertson, C. S. Allen, Y. A. Wu, K. He, J. Olivier, J. Neethling, A. I. Kirkland, and J. H. Warner, Nature communications **3**, 1144 (2012).
[208] K. Kawasumi, Q. Zhang, Y. Segawa, L. T. Scott, and K. Itami, Nature chemistry **5**, 739 (2013).
[209] K. Y. Cheung, X. Xu, and Q. Miao, Journal of the American Chemical Society **137**, 3910 (2015).
[210] F. Barthelat and H. Espinosa, Experimental mechanics **47**, 311 (2007).
[211] H. Gao, B. Ji, I. L. Jäger, E. Arzt, and P. Fratzl, Proceedings of the national Academy of Sciences **100**, 5597 (2003).
[212] B. Ji and H. Gao, Journal of the Mechanics and Physics of Solids **52**, 1963 (2004).
[213] Y. Wang, M. Chen, F. Zhou, and E. Ma, Nature **419**, 912 (2002).
[214] E. Ma, Y. Wang, Q. Lu, M. Sui, L. Lu, and K. Lu, Applied physics letters **85**, 4932 (2004).
[215] N. D. Kim *et al.*, Nano letters **16**, 1287 (2016).
[216] B. Li and M. Arroyo, arXiv preprint arXiv:1703.09371 (2017).
[217] E. Folias, Engineering fracture mechanics **2**, 151 (1970).
[218] E. Folias, International Journal of Fracture Mechanics **5**, 327 (1969).
[219] T. Langer, J. Baringhaus, H. Pfnür, H. Schumacher, and C. Tegenkamp, New Journal of Physics **12**, 033017 (2010).
[220] D. Smirnova, S. H. Mousavi, Z. Wang, Y. S. Kivshar, and A. B. Khanikaev, ACS Photonics **3**, 875 (2016).
[221] M. A. Bissett, M. Tsuji, and H. Ago, Physical Chemistry Chemical Physics **16**, 11124 (2014).
[222] F. Guinea, Solid State Communications **152**, 1437 (2012).
[223] G. F. Schneider, S. W. Kowalczyk, V. E. Calado, G. Pandraud, H. W. Zandbergen, L. M. Vandersypen, and C. Dekker, Nano letters **10**, 3163 (2010).
[224] M. Kothari, M.-H. Cha, and K.-S. Kim, in *APS Meeting Abstracts*2017).
[225] A. N. Kolmogorov and V. H. Crespi, Physical Review B **71**, 235415 (2005).

[226] L. Qiu, J. Z. Liu, S. L. Chang, Y. Wu, and D. Li, Nature communications **3**, 1241 (2012).
[227] Y. Liu, F. Xu, Z. Zhang, E. S. Penev, and B. I. Yakobson, Nano letters **14**, 6782 (2014).
[228] A. Azizi *et al.*, Nature communications **5**, 4867 (2014).
[229] Y. Guo, S. Zhou, J. Zhang, Y. Bai, and J. Zhao, 2D Materials **3**, 025008 (2016).
[230] Y.-C. Lin *et al.*, Nature communications **6**, 6736 (2015).
[231] D. Taha, S. Mkhonta, K. Elder, and Z.-F. Huang, Physical review letters **118**, 255501 (2017).
[232] X. Mu, Z. Song, Y. Wang, Z. Xu, D. B. Go, and T. Luo, Carbon **108**, 318 (2016).
[233] J. Y. Lee, J.-H. Shin, G.-H. Lee, and C.-H. Lee, Nanomaterials **6**, 193 (2016).